\documentclass[a4paper,floatfix,amsmath,amssymb,prb,showpacs,12pt]{revtex4-1}
\usepackage{amsmath}
\usepackage{graphicx}
\pdfoutput=1
\usepackage{subfigure}
\usepackage[pdftex]{hyperref}
\usepackage{epsf}
\usepackage{dcolumn}
\usepackage{bm}
\usepackage{color}
\begin{document}
\newcommand{\newc}{\newcommand}
\newc{\mbf}{\mathbf}
\newc{\boma}{\boldmath}
\newc{\phihat}{\mbox{\boldmath{$\hat{\phi}$}}}
\newc{\thetahat}{\mbox{\boldmath{$\hat{\theta}$}}}
\newc{\beq}{\begin{equation}}
\newc{\eeq}{\end{equation}}
\newc{\beqar}{\begin{eqnarray}}
\newc{\eeqar}{\end{eqnarray}}
\newc{\beqa}{\begin{eqnarray*}}
\newc{\eeqa}{\end{eqnarray*}}
\newc{\bd}{\begin{displaymath}}
\newc{\ed}{\end{displaymath}}

\title{Bead on a rotating circular hoop: a simple yet feature-rich 
dynamical system}
\author{Shovan Dutta$^1$ and Subhankar Ray$^2$}
\affiliation{$^1$ Department of Electronics and Telecommunication Engineering, Jadavpur University, Calcutta 700 032, India}
\affiliation{$^2$ Department of Physics, Jadavpur University, 
Calcutta 700 032, India}

\begin{abstract}
The motion of a bead on a rotating circular
hoop is investigated using elementary calculus and simple symmetry 
arguments. The peculiar trajectories of the bead at different speeds of
rotation of the hoop are presented. Phase portraits and nature of fixed points are studied. Bifurcation is observed with change in the rotational speed of the hoop. At a critical speed of rotation of the hoop, there 
appears an interesting relation between the time period and 
amplitude of oscillation of the bead.
The study introduces several important aspects of nonlinear dynamics.
It is suitable for students having basic understanding in elementary calculus and classical mechanics.
\end{abstract}

\maketitle
\section{Introduction}
\label{intro}
A bead moving on a rotating circular hoop is a classic example studied in 
several textbooks on classical
mechanics\cite{goldstein,jordansmith,strogatz} and nonlinear dynamics.
It exhibits various modes of motion, including some peculiar ones,
such as, oscillations confined to one side of the hoop and 
complete revolutions, for appropriate initial conditions.
It also shows a wide array of
features of dynamical systems. It is useful for demonstrating
different classes of fixed points, bifurcations,
reversibility, symmetry breaking, critical slowing down, trapping 
regions, homoclinic and heteroclinic orbits and Lyapunov functions.
It is also an interesting example of a constrained system, illustrating the use of Lagrange multipliers for determining constraint forces.

In this article we investigate in detail, the dynamics of 
this bead-hoop system. 
The number and nature of equilibrium points alter with change in speed of rotation of the hoop.
This results in some extraordinary modes of motion, phase 
trajectories and bifurcations. At a critical speed of rotation of the
hoop, an interesting relation between the time period and amplitude
of oscillation of the bead is observed.

Our study is based on elementary calculus and simple symmetry 
arguments. A student reader with a background of classical mechanics 
and basic calculus will find the study both accessible and interesting.

The article is organized as follows. In section \ref{physys}, the physical 
system is described. The equation of motion of the system is derived from the
Lagrangian. In section \ref{trajectory}, the effective potential is 
analyzed to find the equilibrium positions of the bead for all possible 
speeds of rotation of the hoop. 
The distinct modes of motion of the bead for different 
initial conditions are described and numerical plots of its trajectories are 
presented. The constraint forces are determined from the method of Lagrange
multipliers in section \ref{energy}. In section \ref{phasebifur}, the nature
of the fixed points of the system are analyzed from symmetry
properties of the system. The phase trajectories and bifurcations are
examined. Finally, we conclude with a discussion on connections to other
systems in section \ref{similar}.

\section{The physical system}
\label{physys}
\begin{figure}[h]
\centering
{\includegraphics[width=5cm]{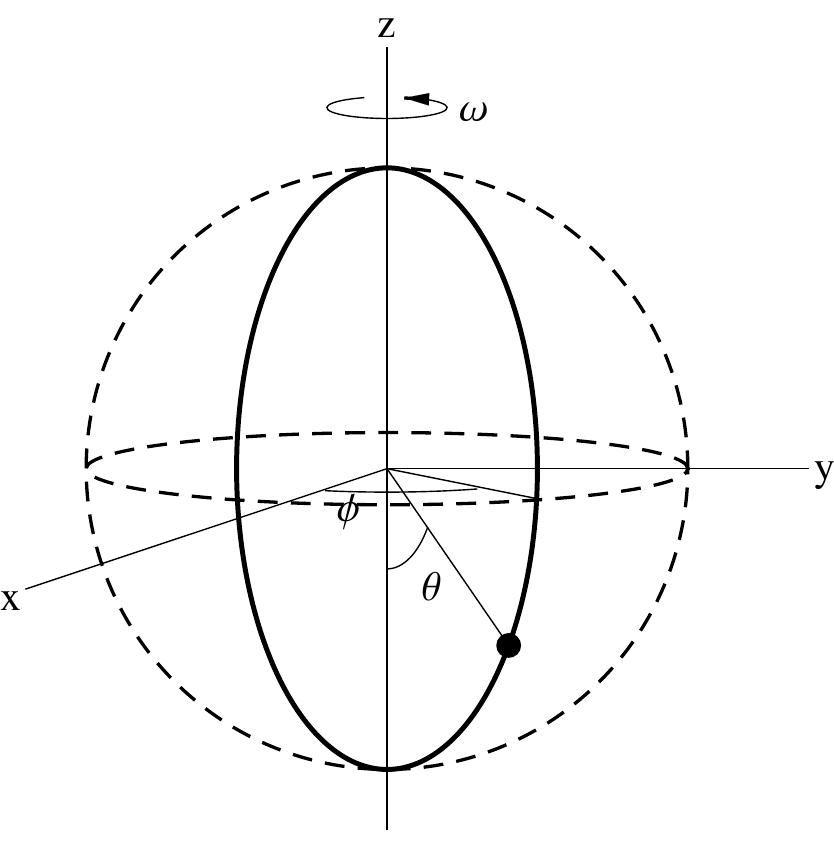}}
\caption{Schematic diagram for bead on the hoop} 
\label{bead_hoop}
\end{figure}

A bead of mass $m$, moves without friction on a circular hoop of 
radius $a$. The hoop rotates
about its vertical diameter with a constant angular velocity $\omega$. 
The position of the bead on the hoop is given by angle $\theta$, 
measured from the vertically downward direction ($-z$ axis),
and $\phi$ is the angular displacement of the hoop from its initial
position on the $x$-axis ( Figure \ref{bead_hoop}).

The kinetic and potential energies of the bead are given by,
\beq \label{kepe}
T= \frac{1}{2} m a^2 ( \dot{\theta}^2  + \sin^2 {\theta}  \dot{\phi}^2 ) \; 
\;\; {\mbox{and}} \;\; V = -mga \cos{\theta} \; ,
\eeq
respectively, where $g$ is the magnitude of the acceleration due to gravity.
The Lagrangian of the system is,
\beq \label{lag1} 
L(\theta, \dot{\theta}) = \frac{ma^2}{2} (\dot{\theta}^2 +\omega^2 \sin^2 \theta )+ m g a \cos \theta 
\;, \; \; \mbox{where} \; \omega = \dot{\phi} \;\; \mbox{(a constant)}.
\eeq
Using the Euler-Lagrange equation, the equation of motion is obtained as,
\beq 
\label{lag2}
\ddot{\theta} = \sin{\theta} (\omega^2 \cos \theta -g/a) .
\eeq
Denoting $\omega^2_c= g/a$, $k = \omega^2 /\omega^2_c$ and defining
$\tau = \omega_c \; t$, the equation of motion may be written in dimensionless form as,
\beq \label{eom}
{\theta}^{''} = \sin{\theta} (k \cos \theta -1) \; ,
\;\; \mbox{where} \;{\theta}^{''}=\frac{d ^2 \theta}{d \tau ^2} .
\eeq
The symmetry of the system about the vertical axis manifests in the invariance of the equation of motion when $\theta$ is changed to $-\theta$.
The system also exhibits time reversal symmetry as (\ref{eom})
is invariant under the transformation $\tau \to -\tau$.

Let us denote $v={\theta}^{'}$ and write (\ref{eom}) as,
\beq
v \frac{d v}{d \theta} = -\sin \theta + k \sin \theta \cos \theta,
\eeq
which upon integration yields,
\beq
\frac{1}{2} v^2 = \cos \theta -\frac{k}{4} \cos 2 \theta + A,
\eeq
where $A$ is a constant of integration.
We can identify a conserved quantity, $\varepsilon$ as,
\beq 
\label{effe}
\varepsilon = \frac{1}{2}{\theta ^{'}}^2 - \frac{1}{2}k \sin^2 \theta -
\cos \theta \; .
\eeq
It may be termed as the effective energy that corresponds to the
system described by (\ref{eom}).
However, it is to be noted that, the mechanical energy of the original bead-hoop system
is not conserved due to the work done in preserving the constant
angular velocity of the hoop.
Equation (\ref{effe}) provides a natural separation of the effective energy
into effective kinetic energy $ \frac{1}{2}{\theta^{'}}^2 $ and 
effective potential energy, $U(\theta)$, given by,
\beq \label{effpot}
U(\theta) = - \frac{1}{2}k \sin^2 \theta - \cos \theta \; .
\eeq
As $U(\theta)$ is an even function of $\theta$, 
we restrict our attention to $ \theta \in [0,\pi]$.
${\theta^{'}}^2$ is completely determined by $\theta$ and $\varepsilon$.
The sign of $\theta^{'}$, however, cannot be determined uniquely
from (\ref{effe}).
The parameter $k$ grows with increasing angular velocity $\omega$ of the hoop
and $k=0$ implies that the hoop is stationary. 

It is useful to consider the first and second derivatives of $U(\theta)$, 
\beqar
\label{derv1}
U^{'}(\theta) &=& \frac{d U}{d \theta}= - \sin \theta
(k \cos \theta -  1) \; , \;\; {\mbox{and}} \\
U^{''}(\theta) &=& \frac{d^2 U}{d \theta^2} = - 
(k \cos 2\theta -  \cos \theta) \; .
\label{derv2}
\eeqar
Comparison of (\ref{eom}) and (\ref{derv1}) shows that,
\beq
\label{theta_acc}
\theta^{''} = -U^{'}(\theta)=- \sin\theta (1 - k \cos\theta) \, .
\eeq
\begin{figure}[h]
\centering
{\includegraphics[width=8 cm,height = 5.5cm]{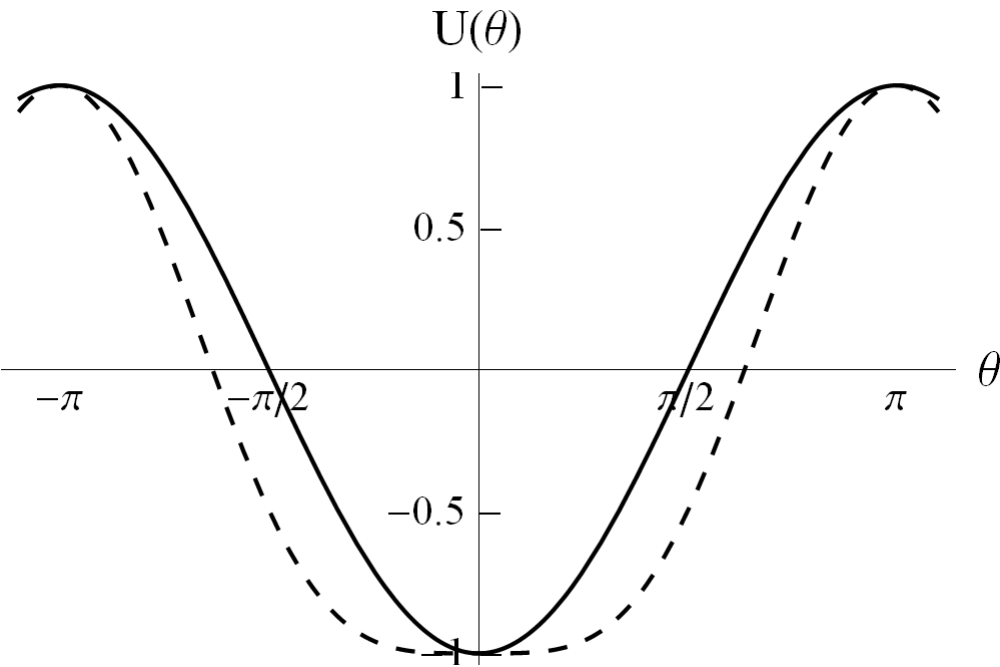}}
\caption{Minimum of $U(\theta)$ is flatter and maxima sharper as
$k \to 1$. The solid curve is for $k = 0.010$, dashed curve is for $k = 0.998$} 
\label{epotk}
\end{figure}
The effective potential has two extrema at 
$\theta =0$ 
and $\theta = \pi$, where the first derivative of the potential vanishes.
Evaluating the second derivative at these points, one finds that
$\theta=0$ is a minimum ($U^{''}(0) >0$) and $\theta= \pi$ is a maximum
($U^{''}(0) <0$) of the effective potential as shown 
in Figure \ref{epotk}. Due to the right-left symmetry of the hoop, one concludes, $\theta = -\pi$ is also a maximum point of the effective potential.
At these points the values of the effective potential are 
$U(0)=U_{min}=-1$ and $U(\pm \pi)=U_{max}=1$.

\section{Modes of motion and trajectories of the bead}
\label{trajectory}
\subsection{$\mbf{0 \le k < 1}$}
When $k=0$, the effective potential takes the form of the potential energy of
a pendulum. As $k$ increases from zero towards $1$, 
$U^{''}( 0) \to 0$ and $|U^{''}(\pm \pi)| \to 2$, implying that the minimum at
$\theta = 0$ is flatter and the maxima at $\theta = \pm \pi$ 
are sharper (Figure \ref{epotk}).

For $k \neq 0$, at different effective 
energy values, the actual motion of the bead will depend on the initial 
conditions which we discuss below.\\
\begin{figure}[h]
\centering
{\includegraphics[width=8cm,height =5.5cm]{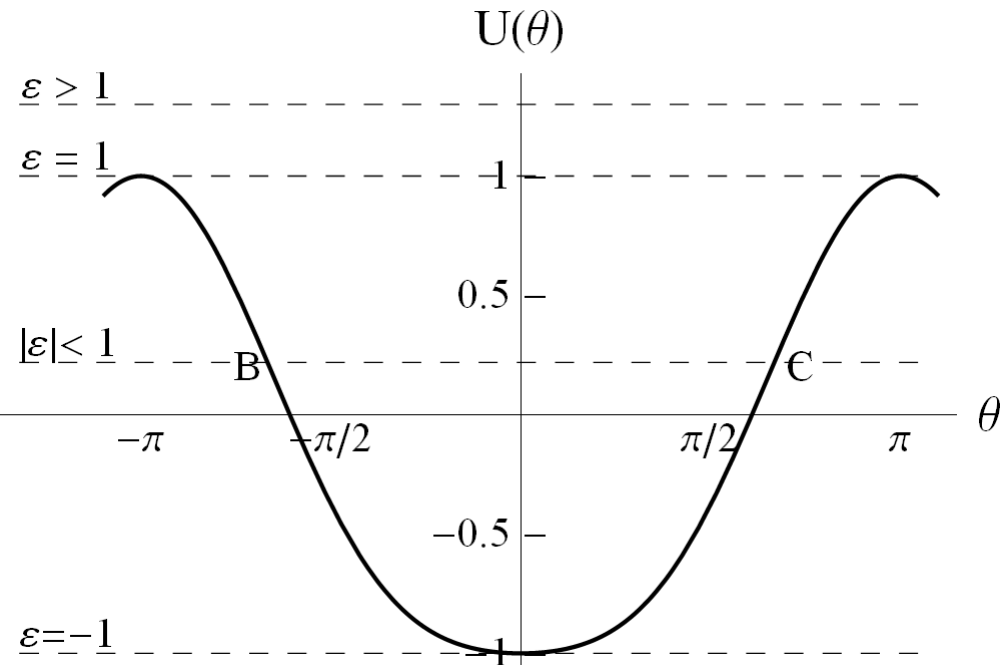}}
\caption{Effective potential $U(\theta)$ versus $\theta$ for $0 \leq k<1$}
\label{epot1}
\end{figure}

\noindent
(i) Effective energy, $\varepsilon = -1$ \\
${\theta^{'}}^2 =2(\varepsilon - U(\theta)) $ must be positive, hence the 
lower bound of $\varepsilon$ is equal to the minimum value of $U(\theta)$ 
which is $-1$ (Figure \ref{epot1}). This corresponds to the initial 
condition $\theta(0)=0$ and $\theta^{'}(0)=0$. From (\ref{theta_acc}) 
however, $\theta^{''} = 0$ 
as well, independent of the value of $k$.
Hence, the bead stays at the bottom of the hoop ($\theta=0$) for all times. \\

\begin{figure}[h]
\centering
\subfigure[$\, \tau = 28$]{\includegraphics[width=1.8in]
{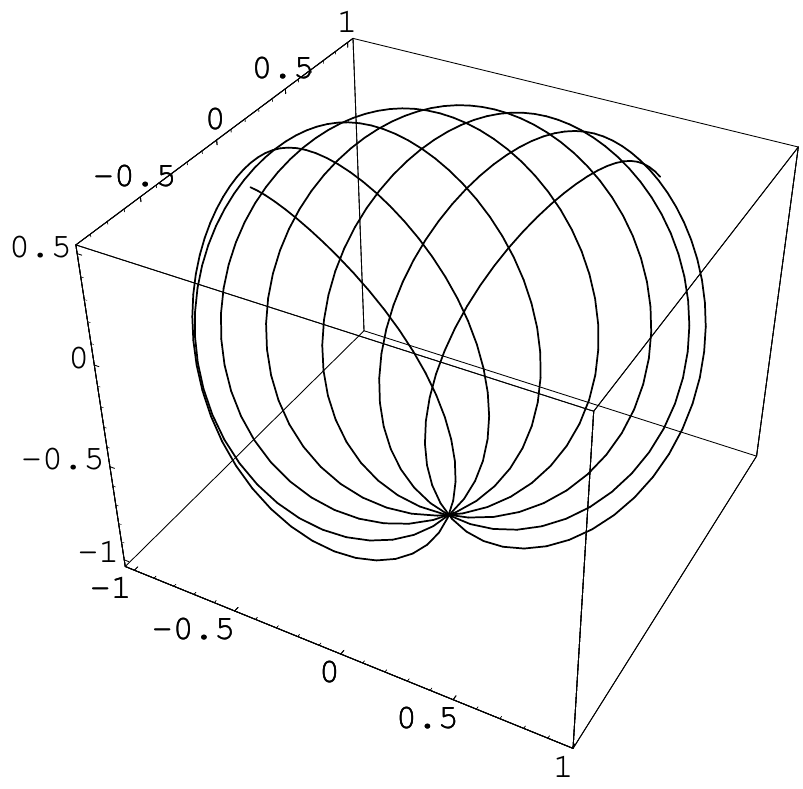}}
\subfigure[$\, \tau = 62.5$]{\includegraphics[width=1.8in]
{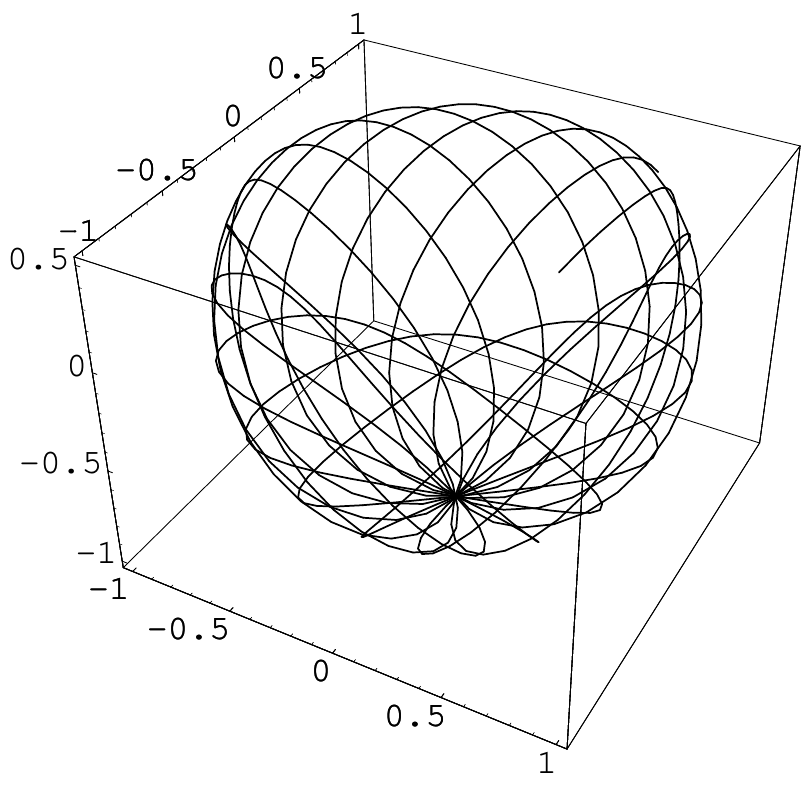}}
\subfigure[$\, \tau = 62.5$, top view]{\includegraphics[width=1.8in]
{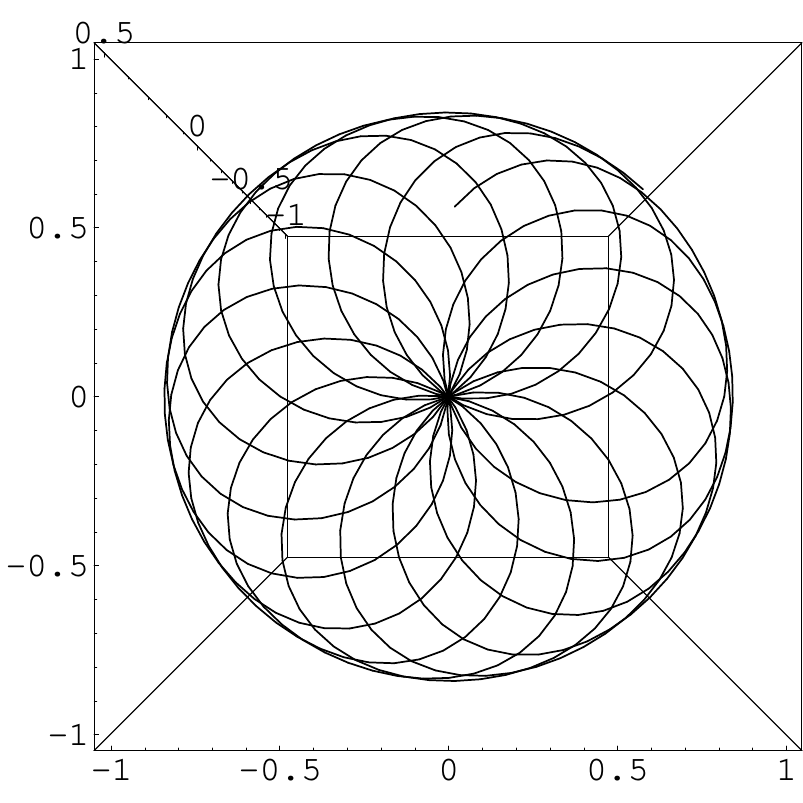} }
\caption{Trajectory for k=0.75, $\theta(0)=2\pi/3$, $\theta^{'}(0)=0$ 
at different $\tau$ values}
\label{traj1}
\end{figure}

\noindent
(ii) $-1 < \varepsilon  < 1$ \\
The angular position $\theta$ of the bead oscillates between $-\theta_0$ and $+\theta_0$, corresponding to the turning points $B$ and $C$ 
shown in Figure \ref{epot1}. At these points $\theta^{'}=0$, and 
the effective potential equals the effective energy.
\begin{figure}[h]
\centering
{\includegraphics[height=10cm]{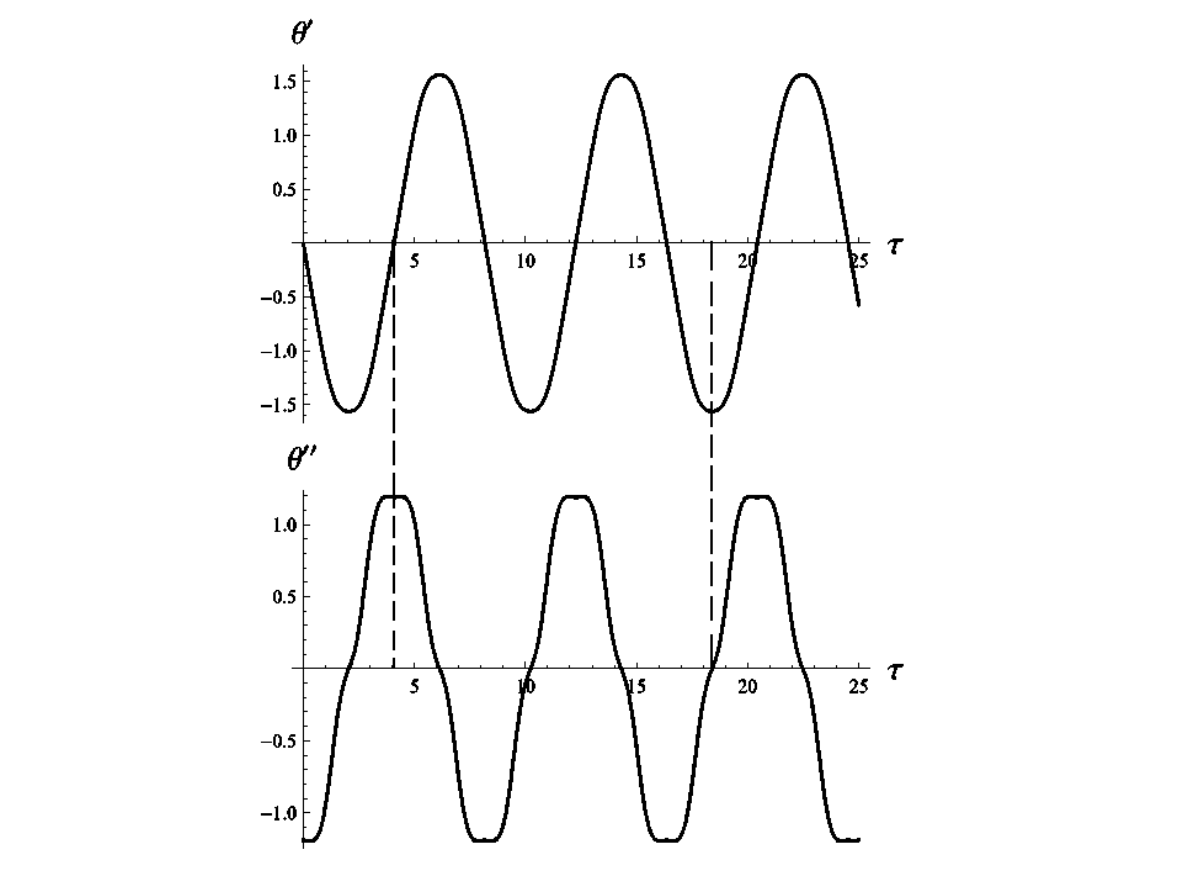}}
\caption{Out of phase oscillation of $\theta^{'}$ and $\theta^{''}$ for $k=0.75$, $\theta_0 = 2\pi/3$} 
\label{outphase}
\end{figure}
From (\ref{eom}), $\theta^{''} < 0$ and this restoring acceleration ensures that $\theta$ approaches 0. At $\theta=0$, acceleration $\theta^{''}$ is zero, whereas the velocity $\theta^{'}$ attains 
its maximum magnitude. Hence, the bead slides past the 
bottom of the hoop and $\theta$ decreases further, becoming negative. When $\theta <0$, 
$\theta^{''}$ becomes positive,
and this acceleration reduces the magnitude of $\theta^{'}$, making it zero at $-\theta_0$.
$\theta^{''}$ remains nonzero and positive at $-\theta_0$. The acceleration $\theta^{''}$ is positive for all negative values of $\theta$. At $\theta = -\theta_0$, $\theta^{''} > 0$ and velocity is momentarily zero. From the next instant, velocity begins to increase as $\theta^{''} > 0$.
Thus $\theta^{'}$ and $\theta^{''}$ oscillate out of phase with each
other as shown in Figure \ref{outphase}.
At $-\theta_0$, $\theta^{''} > 0$, $\theta^{'}$ becomes positive, $\theta$ begins to
increase. The bead retraces its path reaching $\theta_0$ and the process
is repeated.
The amplitude of oscillation increases with $\varepsilon$.
The time period of the oscillation of the bead between $\pm \theta_0$ 
is given by,
\beq 
\label{time}
T = 4 \int_0^{\theta_0} \frac{d \, \theta}{\sqrt{(\cos \theta - 
\cos \theta_0)(2-k(\cos \theta + \cos \theta_0))}} \; .
\eeq
Figure \ref{traj1} shows the trajectory of the bead after the elapse of
different time intervals, with $k = 3/4$ and
amplitude $\theta_0 = 2\pi/3$. 
As the bead goes through half an oscillation (with
amplitude $2\pi/3$ and $k = 3/4$), the hoop rotates through
an angle given by,
\bd
\Delta \phi = \omega \left ( \sqrt{\frac{a}{g}} \, \frac{T}{2} \right )
= \frac{\sqrt{k}}{2}T = 3.539 \; {\mbox{radians}} \; .
\ed

\begin{figure}[h]
\centering
{\includegraphics[height=5cm,width=4.5cm]{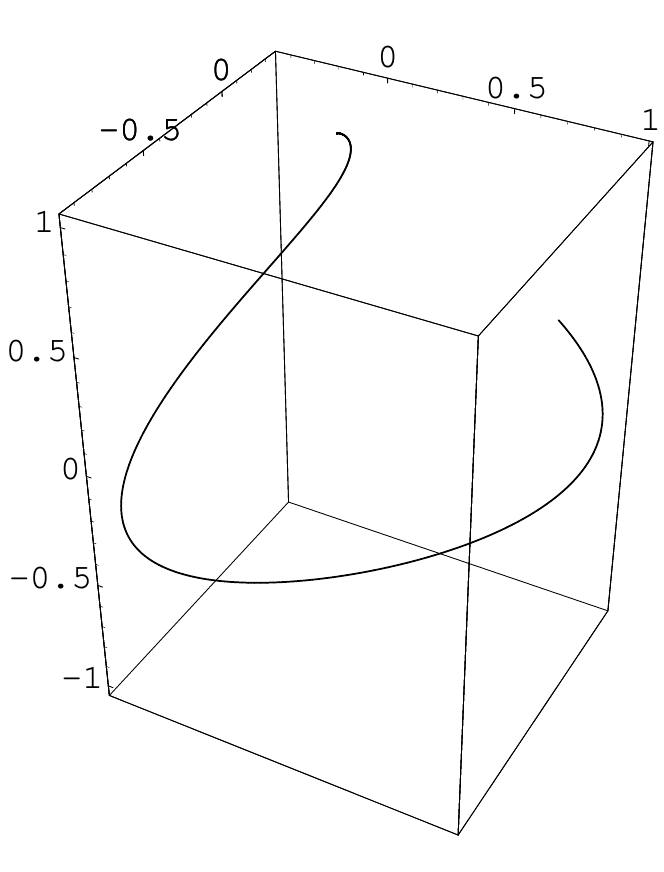}}
\caption{Trajectory for $k=0.75$, $\theta(0)=2\pi/3$ and 
$\varepsilon = 1 $} 
\label{approach_pi}
\end{figure}

\noindent
(iii) $\varepsilon = 1$ \\
At $\theta = \pm \pi$, the effective potential energy is also $+1$. Thus the effective kinetic energy and $\theta^{'}$ is zero. From the expression of acceleration in (\ref{eom}), we find $\theta^{''} = 0$ at $\theta = \pm \pi$. This agrees with the fact that $\theta = \pm \pi$ are the position of maximum potential energy. From (\ref{effe}), we obtain,
\beq
{\theta^{'}}^2 = 4 \cos^2\frac{\theta}{2} \, [ 1 + k \sin^2 
\frac{\theta}{2}] \; .
\eeq
The time taken to reach the highest point ($\theta = \pi$) from an initial point $\theta_0$ may be computed as,
\beq
T = \frac{1}{2}  \int_{\theta_0}^{\pi} \frac{d \, \theta}{\cos{(\theta/2)}\sqrt{1 + k \sin^2{(\theta/2)}}}
\eeq
The integral can be transformed with the substitution $u = \pi - \theta$, as,
\beq
T = \frac{1}{2} \int_{0}^{\pi - \theta_0} \frac{d \, u}{\sin{(u/2)}\sqrt{1 + k \cos^2{(u/2)}}}
\eeq
In the neighbourhood of $u = 0$, corresponding to 
$\theta$ close to $\pi$, the integrand behaves as $\frac{(1 + k)^{-1/2}}{u}$ 
and hence, the integral diverges. This implies that the bead approaches 
$\theta = \pi$ infinitesimally slowly and takes an infinite 
amount of time to reach the top of the hoop (Figure \ref{approach_pi}).
By virtue of time-reversal symmetry and the left-right symmetry of the hoop, 
for $\theta^{'} (0) < 0$, the bead reaches the bottom where $\theta^{'}$ 
has the maximum magnitude of 2 (from (\ref{effe})) and 
then asymptotically approaches $\theta = -\pi$ at ever-decreasing speed. \\

\begin{figure}[h]
\centering
\subfigure[$\, \tau = 20$]{\includegraphics[width=1.8in]
{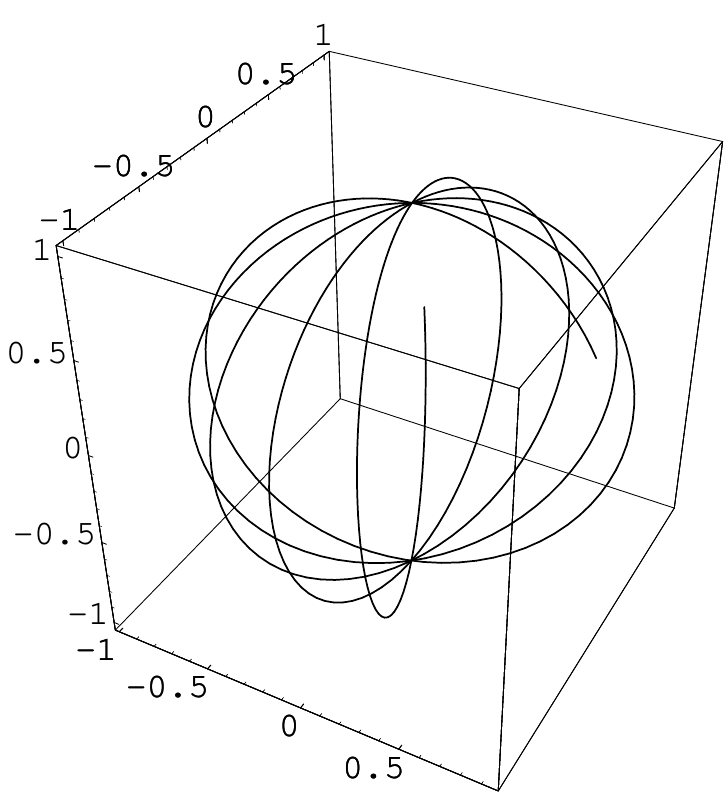}}
\subfigure[$\, \tau = 40$]{\includegraphics[width=1.8in]
{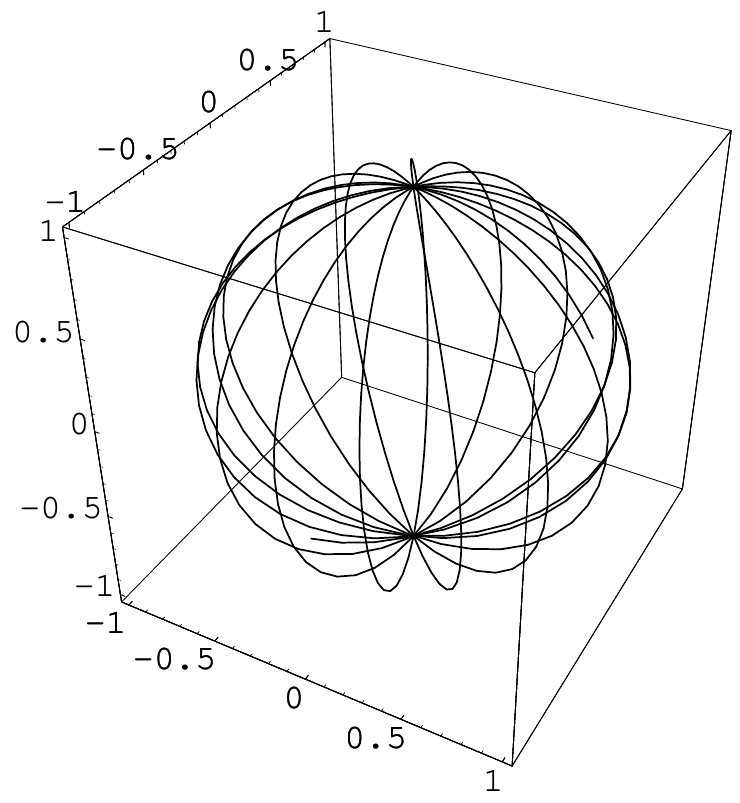}}
\subfigure[$\, \tau = 200$]{\includegraphics[width=1.8in]
{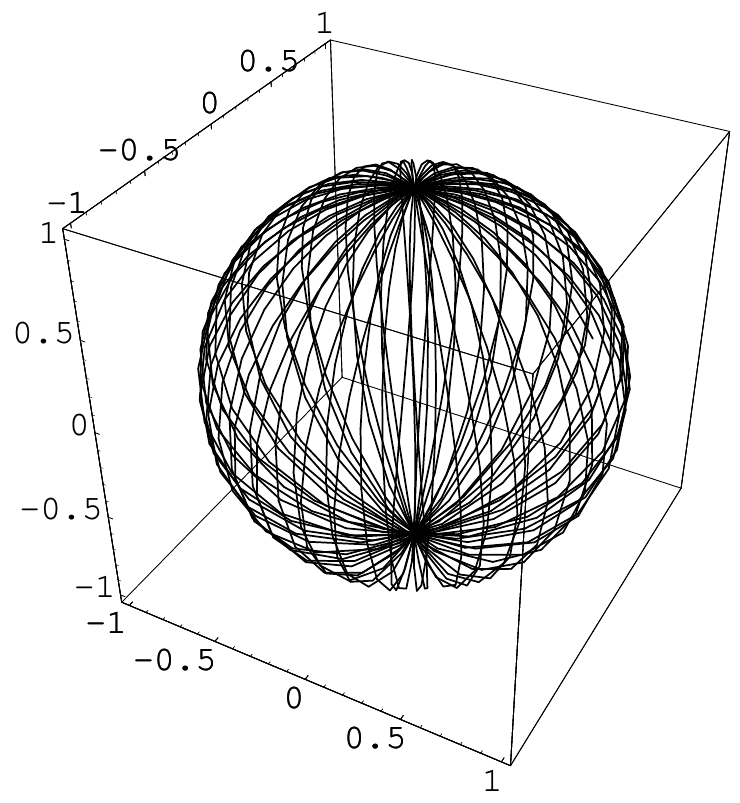} }
\caption{Trajectory for k=0.01, $\theta(0)=2\pi/3$, $\theta^{'}(0)=1.4$ 
for different $\tau$.}
\label{full1}
\end{figure}

\noindent
(iv) $\varepsilon > 1$ \\
As the maximum value of $U(\theta) = 1$, $\varepsilon -U(\theta) > 0$ and hence
$\theta^{'}$ is not zero for any $\theta$.
Thus if the bead starts initially with $\theta^{'}(0) > 0$ and at a
position $\theta(0) \in (0, \pi)$, then $\theta^{''}(0) <0$ and $\theta^{'}$ 
starts to decrease. $\theta^{'}$ attains its minimum value when the 
bead reaches $\theta = \pi$. 
At this position, $\theta^{''} =0$, however, $\theta^{'}$ is not zero and 
$\theta$ continues to increase. Hence the bead moves past the top 
of the loop. For $\theta \; \in (\pi, \frac{3 \pi}{2})$, 
$\theta^{''}$ is positive and $\theta^{'}$ increases, becoming maximum 
when the bead reaches the bottom of the hoop. 
Thus $\theta$ increases or decreases indefinitely, 
depending on the sign of $\theta^{'}(0)$ and the bead slides over the entire
hoop periodically (Figure \ref{full1}). For $\theta^{'}(0) <0$,
the same rotation occurs 
in the opposite sense. The maximum and minimum values of 
$\theta^{'}$ can be calculated from the expression of the effective energy.
The time period of rotation is given by,
\beq
\label{trot}
T = 2 \int_{0}^{\pi} \frac{d \, \theta}{\sqrt{(\theta^{'}(0))^2 - (\cos\theta - \cos\theta_0)[2 - k(\cos\theta + \cos\theta_0)]}}
\eeq
When $|\theta^{'}(0)|$ is very large, nearly all the energy
of the bead is the kinetic energy of rotation. Hence the influence 
of gravity becomes small in comparison, and this rotation may be 
considered to be uniform. In this limit, the time period approaches 
the value $2\pi /|\theta^{'}(0)|$.
\subsection{$\mbf{k = 1}$}
$U(\theta)$ is maximum at $\theta = \pi$ as before.
However, at $\theta =0$, both 
$U^{'}(0)$ and $U^{''}(0)$ vanish.
On expanding $U(\theta)$ about $\theta =0$, we get,
\bd
U(\theta) = -1 + \frac{\theta^4}{8} + {\mbox{O}}(\theta^6) \; .
\ed
For $ |\theta| \ge 0$, in the neighbourhood of $\theta =0$, $U(\theta)$ 
is always greater than $U(0)$.
Hence $\theta =0$ is a minimum of the potential. The potential is more 
flat than for the case $k < 1$.
Depending on the values of $\varepsilon$, the bead can undergo all the types 
of motion described in the previous subsection.
However, for oscillatory motion,
the period of oscillation for a given amplitude is larger 
when $k=1$ than when $k <1$. 
This becomes clear if (\ref{eom}) is expanded 
in the vicinity of $\theta = 0$,
\beqa
\theta^{''} &= (\theta - \frac{\theta^3}{3!} + ...)[k(1 - \frac{\theta^2}{2!} + ...) - 1] \\
&= -(1 - k) \theta - \big( \frac{2k}{3} - \frac{1}{6}
\big) \theta^3 + {\mbox{O}}(\theta^5) 
\eeqa
When $k < 1$, $\theta^{''} \sim {\mbox{O}}(\theta)$, however, for $k=1$, 
$\theta^{''} \sim {\mbox{O}}(\theta^3)$. For small amplitude $\theta$ 
($\theta \to 0$),
$\theta^{''}_{k=1} (\sim {\mbox{O}}(\theta^3))$ is much smaller than 
$\theta^{''}_{k<1} (\sim {\mbox{O}}(\theta))$. Thus $\theta^{'}$ 
changes much more slowly 
in the neighbourhood of $\theta = 0$ for $k=1$. 
Consequently, the period of oscillation $T$, 
for the same amplitude is significantly larger.
\begin{table}[ht]
\centering
\caption{Time period of oscillations for different amplitudes 
at various $k$ values.}
\vspace{0.2cm}
\begin{tabular}{|l|c|c|c|c|}
\hline
$T$ about $\theta = 0$ & k = 0 & k = 0.5 & k = 0.75 & k = 1 \\
\hline
amplitude $\pi/10$ & 6.32 & 8.78 & 12.00 & 33.71 \\
\hline
amplitude = $\pi/20$ & 6.29 & 8.86 & 12.41 & 66.93 \\
\hline
amplitude = $\pi/40$ & 6.29 & 8.88 & 12.53 & 133.62 \\
\hline
amplitude = $\pi/80$ & 6.28 & 8.88 & 12.55 & 267.12 \\
\hline
\end{tabular}
\label{largeT}
\end{table} 

\begin{figure}[h]
\centering
{\includegraphics[width=10cm]{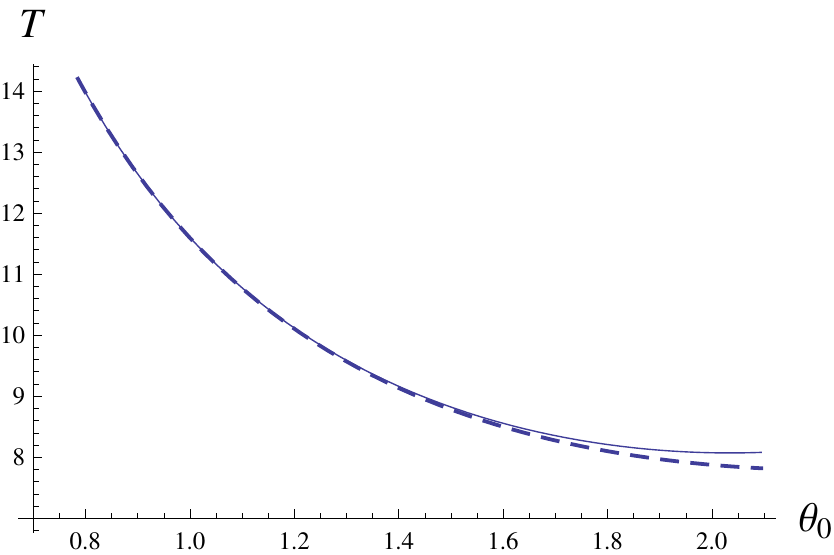}}
\caption{Time period vs amplitude: solid curve represents accurate 
plot given by (\ref{inv_amp}), dashed curve represents approximate 
plot given by (\ref{approximate_T})}
\label{timeplot}
\end{figure}

\begin{figure}[h]
\centering
{\includegraphics[width=10cm]{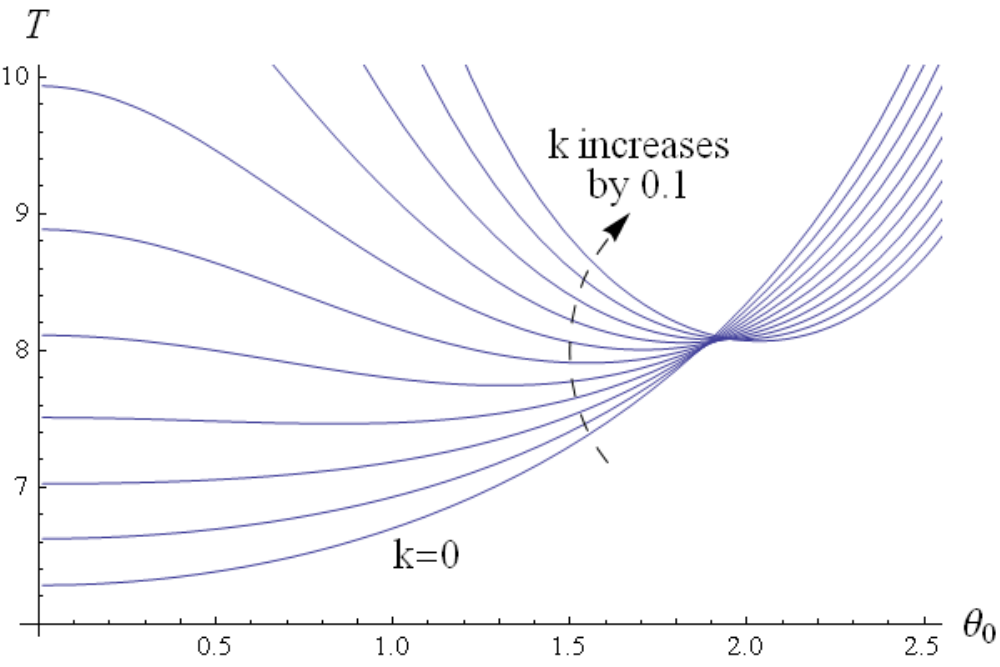}}
\caption{Variation of time period with amplitude at various values of k}
\label{tvar}
\end{figure}

Table \ref{largeT} lists the time periods computed for 
different amplitudes at various values of $k$. The values in the table reveal
a very interesting pattern. For small amplitudes, the time period is essentially 
independent of amplitude when $0 \leq k < 1$; but for $k=1$, the time period increases inversely as the decrease in amplitude. This is explained as follows. 
For $0 \leq k < 1$, the effective potential is approximately parabolic in 
the neighbourhood of $\theta = 0$, analogous to the simple pendulum. Thus, 
the small amplitude motion of the bead is simple harmonic, for 
which the time period is independent of amplitude. However, this parabolic 
approximation does not hold when $k=1$, as the second derivative of the 
effective potential, $U^{''}(\theta)$, vanishes at $\theta = 0$. 
Using (\ref{time}), the time period of oscillation for $k=1$ can be 
calculated as follows,
\beq
\label{inv_amp}
T = 4 \int_0^{\theta_0} \frac{d \, \theta}{\sqrt{(\cos \theta - 
\cos \theta_0)(2-\cos \theta - \cos \theta_0)}} \; .
\eeq
Expanding the integrand for small $\theta_0$ gives,
\beq
\nonumber T = 4  \int_0^{\theta_0}\frac{d \, \theta}{\sqrt{(\frac{\theta_0^2 - \theta^2}{2!} - \frac{\theta_0^4 - \theta^4}{4!} + \frac{\theta_0^6 - \theta^6}{6!} + \dots)(\frac{\theta_0^2 + \theta^2}{2!} - \frac{\theta_0^4 + \theta^4}{4!} + \frac{\theta_0^6 + \theta^6}{6!} + \dots)}}
\eeq
Retaining terms up to eighth order in $\theta_0$, 
\beqar
\nonumber T &=& 4 \int_0^{\theta_0} \frac{d \, \theta}{\sqrt{\frac{1}{4} (\theta_0^4 - \theta^4) - \frac{1}{24} (\theta_0^6 - \theta^6) + \frac{1}{320} (\theta_0^8 - \theta^8)}} \\
\nonumber &=& \frac{8}{\theta_0^2} \int_0^{\theta_0} \frac{d \, \theta}{\sqrt{1 - (\frac{\theta}{\theta_0})^4 - \frac{\theta_0^2}{6} ( 1 - (\frac{\theta}{\theta_0})^6) + \frac{\theta_0^4}{80} ( 1 - (\frac{\theta}{\theta_0})^8)}}
\eeqar
Defining a new variable $y = \theta/\theta_0$, this can be expressed as,
\beqar
\nonumber T &=& \frac{8}{\theta_0} \int_0^{1} \frac{d \, y}{\sqrt{1 - y^4 - \frac{\theta_0^2}{6} ( 1 - y^6) + \frac{\theta_0^4}{80} ( 1 - y^8)}} \\
\nonumber &=& \frac{8}{\theta_0} \int_0^{1} \frac{d \, y}{\sqrt{1 - y^4}}\Bigg[1 - \frac{\theta_0^2}{6} \frac{1 - y^6}{1 - y^4} + \frac{\theta_0^4}{80} \frac{1 - y^8}{1 - y^4}\Bigg]^{-1/2}
\eeqar
Expanding the quantity within brackets in a binomial series and retaining 
terms upto fourth order in $\theta_0$,
\beq
\label{approximate_T}
T = \frac{A}{\theta_0} + B \theta_0 + C \theta_0^3 + O(\theta_0^5)
\eeq
where the constants are given by,
\beqar
A = 8 \int_0^1 \frac{d y}{\sqrt{1 - y^4}} = 10.49 \\
B = \frac{2}{3} \int_0^1 \frac{1 - y^6}{(1 - y^4)^{3/2}} d y = 1.04 \\
C = \int_0^1 \Bigg( \frac{1}{12} \frac{(1 - y^6)^2}{(1 - y^4)^{5/2}} - \frac{1}{20} \frac{1 - y^8}{(1 - y^4)^{3/2}} \Bigg) d y = 0.07
\eeqar
This shows that for small $\theta_0$, $T \approx A/\theta_0$, which 
explains why $T$ doubles when the amplitude is halved. Equation 
(\ref{approximate_T}) gives a very good approximation even if $\theta_0$ 
is not close to $0$, e.g., for $\theta_0 = \pi/2$, the error is 
only $-0.58\%$ (Figure \ref{timeplot}). This can be 
accounted for by noting that the coefficients 
of higher order terms in $\theta_0$ decrease rapidly. Thus the expansion 
converges very fast. As an example, for $\theta_0 = 1$, the third term in 
(\ref{approximate_T}) contributes only about $0.67 \%$ of the first term.
Another point to be noted is that the time period of oscillation for $k=0$,
increases with amplitude, whereas, for higher values of $k$ ($k>0.3$),
the time period decreases initially with amplitude (Fig \ref{tvar}). As 
Eventually for all $k$, the time period approaches $\infty$ as the amplitude 
approaches $\pi$.
\subsection{$\mbf{k > 1}$}
When $k > 1$, in addition to $\theta =0$ and $\theta =\pi$, a new 
extremum of the potential 
appears at $\Omega_1 = \cos^{-1}(1/k)$. From (\ref{derv2}), 
it may be shown,
\beq
U^{''}(0)  = -(k -1)  < 0 \;, \;\;\;
U^{''}(\pi)  = -(k + 1)  < 0 \; , \;\;\;
U^{''}(\Omega_1) = \left( k -\frac{1}{k} \right)  > 0 \; . 
\eeq

\begin{figure}[h]
\centering
{\includegraphics[width=8cm,height =5.5cm]{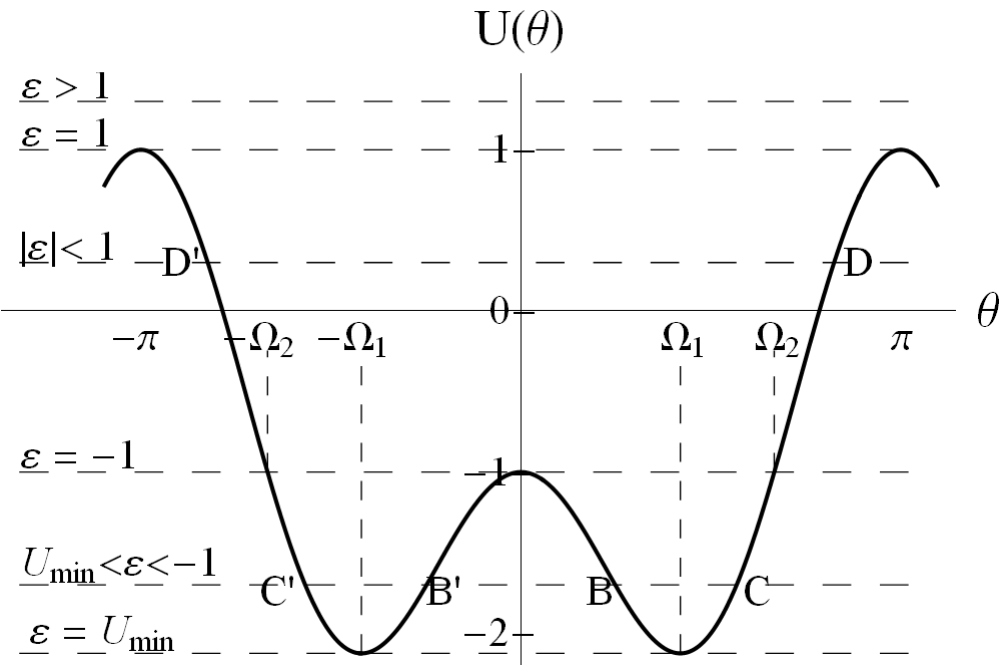}}
\caption{Effective potential $U(\theta)$ versus $\theta$ for $ k > 1$.} 
\label{epot2}
\end{figure}

$\theta=\pi$ remains a maximum of the effective potential, however, 
$\theta =0$ which was a minimum for $k \le 1$, is now a maximum. 
The minimum is at 
$\theta = \Omega_1$, where, 
\bd
U(\Omega_1) = -\frac{1}{2} \left( k + \frac{1}{k} \right) \; .
\ed
The effective potential is now a double well as shown in Figure \ref{epot2}.
Due to left-right symmetry of the hoop, $-\Omega_1$ is also a minimum of
the potential. For $k > 1$, the potential value at $\Omega_1$ is less
than $-1 $, and the value decreases monotonically as 
$k$ increases. So, the minimum at $\theta = \Omega_1$ becomes deeper 
with increase in $k$. As $k$ increases from $1$ towards $\infty$,
$\Omega_1$ increases from $0$ to $\pi/2$ in a monotonic manner.
The maximum at $\theta =0$ is less than the maximum at $\theta=\pi$. \\
As in the previous section, the motion of the bead 
in different effective energy regions can 
be identified as follows. \\

\begin{figure}[h]
\centering
\subfigure[$\, k=2$, $R=4.83$]{\includegraphics[width=2.0in]
{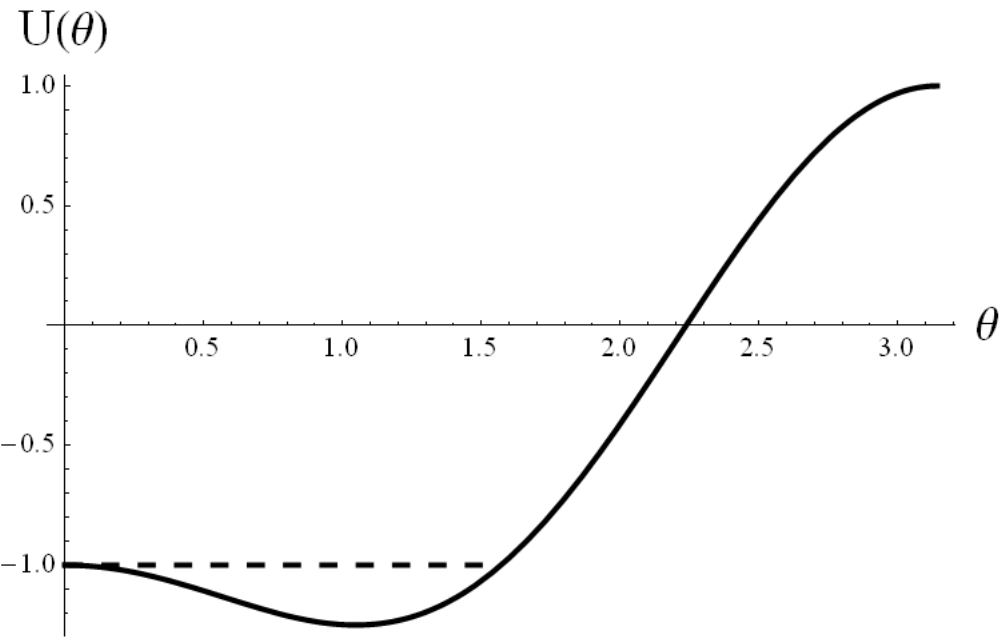}}
\subfigure[$\, k=5$, $R=1.58$]{\includegraphics[width=2.0in]
{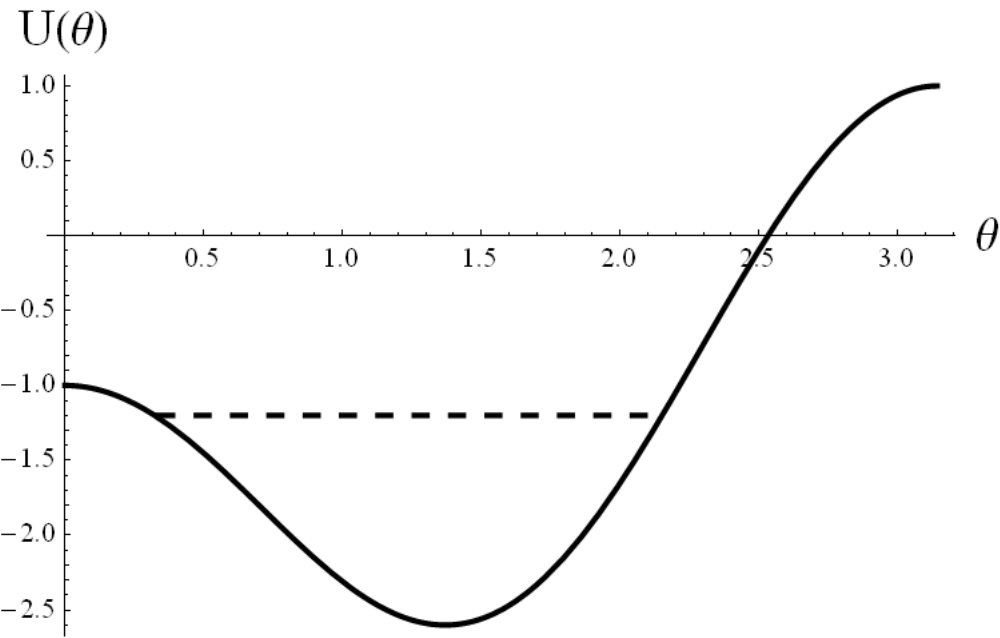}} 
\subfigure[$\, k=20$, $R=1.15$]{\includegraphics[width=2.0in]
{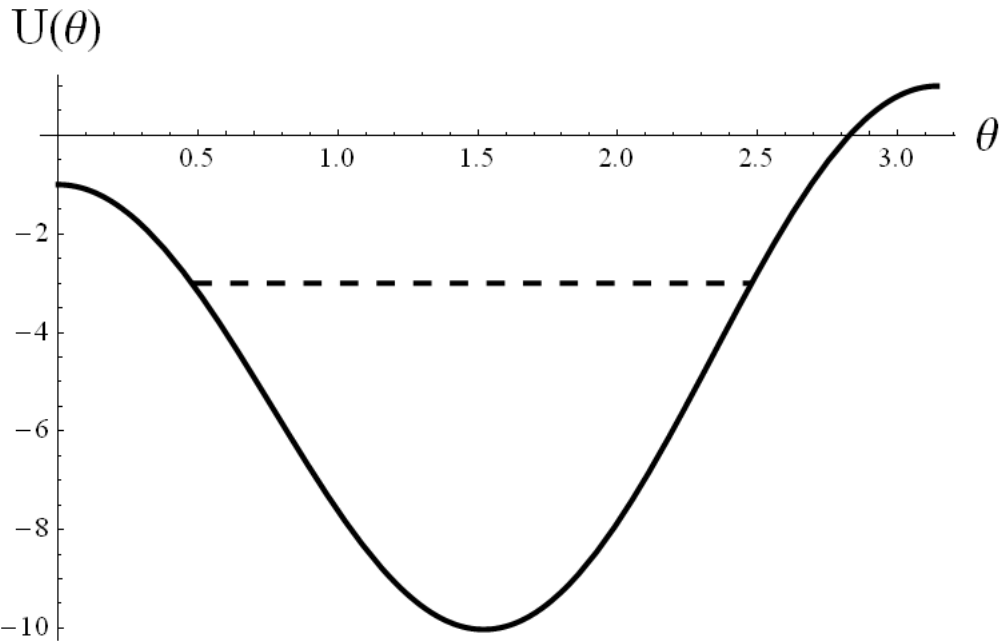} } 
\caption{Asymmetry in the effective potential about its minimum $\Omega_1$
at different $k$ values for $\theta >0$.}
\label{asympot}
\end{figure}

\begin{figure}[h]
\centering
\subfigure[$\, k = 2$]{\includegraphics[width=2.0in]
{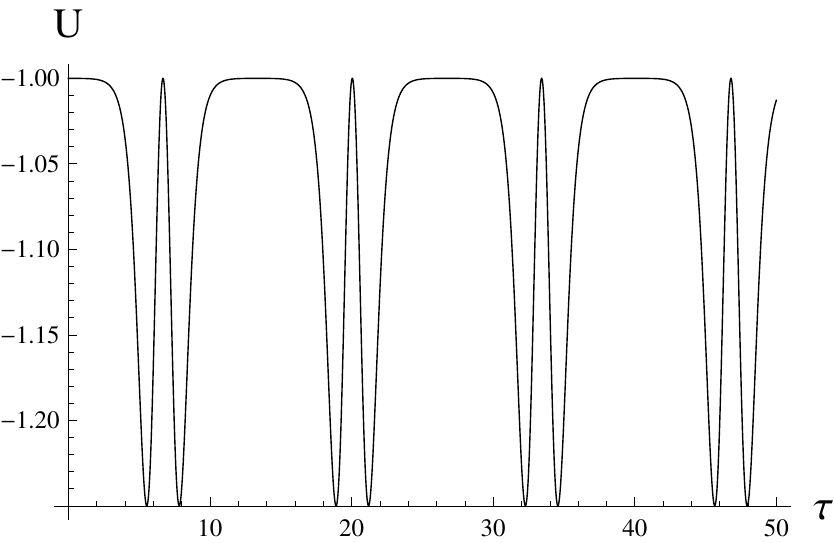}}
\subfigure[$\, k = 5$]{\includegraphics[width=2.0in]
{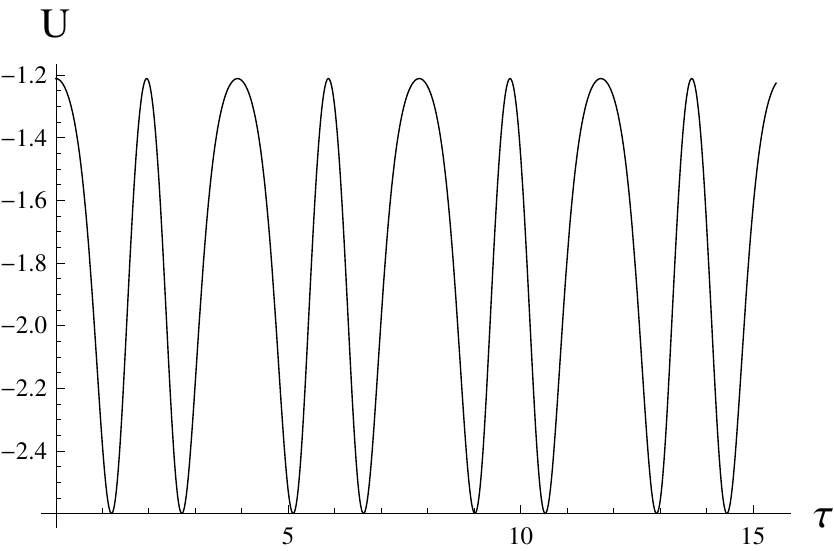} }
\subfigure[$\, k = 20$]{\includegraphics[width=2.0in]
{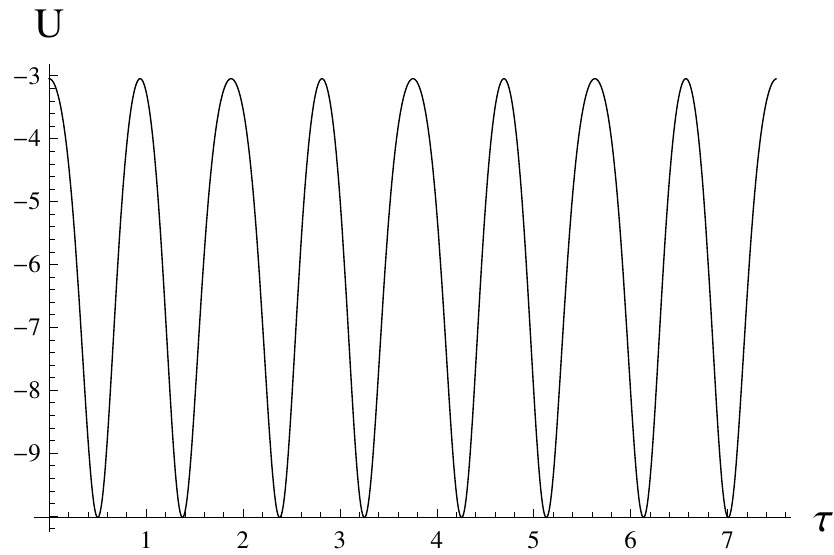} } 
\caption{Change in the potential energy of the bead with time at
different $k$ values.}
\label{asymk}
\end{figure}

\noindent
(i) $\varepsilon = -1/2(k + 1/k)$ : \\
The only two possible positions of the bead are at $ \pm \Omega_1$,
where both $\theta^{'}$ and $\theta^{''}$ are zero.
Here, even though the minima of the effective potential are symmetric; 
depending on the initial conditions, one or the other of these minima
will be chosen, leading to a breaking of symmetry\cite{sivardiere}.
As the hoop rotates about its vertical axis, the bead simply
moves on a circle of radius $ a \sin \Omega_1$. \\
\noindent
(ii) $ -1/2(k + 1/k) < \varepsilon < -1 $ : \\
The bead oscillates between 
$\theta = \pm \alpha$ to $\theta = \pm \beta$
corresponding to the turning points $B$, $C$ about $\Omega_1$,
or $B'$, $C'$ about $- \Omega_1$.
This is similar to the oscillations encountered for the cases
$0 \le k < 1$ and $k=1$, except that, here, the oscillations are 
confined to one side of the hoop (with respect to
the vertical axis) and are not symmetric about the minima $\pm \Omega_1$. 
The bead swings for longer time below the equilibrium point than above it. 
As $\theta^{'}=0$ at $\alpha$ and $\beta$, we may write using (\ref{effe}), 
\beqa
\cos \alpha + \cos \beta = 2 \cos\Omega_1 \;\;\; \mbox{or, } \nonumber\\
\cos\frac{\alpha + \beta}{2} \cos\frac{\alpha - \beta}{2} = 
\cos\Omega_1 \nonumber
\eeqa
Hence, $\cos\frac{\alpha + \beta}{2} > \cos\Omega_1$ or, $\Omega_1 - 
\alpha > \beta - \Omega_1$. Thus the turning point $B$ in Figure
 \ref{epot2} is farther from the point of equilibrium $\Omega_1$,
than the other turning point $C$.
The shape of the effective potential also confirms this (see Figures 
\ref{epot2} and \ref{asympot}).
About $\theta = 0$, the potential is symmetric, whereas, about 
$\theta = \Omega_1$, it has a steeper ascent towards $\pi$ 
than toward $0$. 
This asymmetry in the shape of the potential decreases as $k$ 
assumes larger values (Figure \ref{asympot}). 
This is illustrated in Figure \ref{asympot} where
this asymmetry is shown for $\theta >0$.
In Figure \ref{asymk}, the variation of the potential energy of the
bead as it oscillates about $\Omega_1$ is shown with $\alpha 
\approx \Omega_1 - 60^0$. $R$ is the ratio of the time spent below
the equilibrium position $\Omega_1$, to the time spent above it.
\begin{figure}[h]
\centering
\subfigure[$\, \tau = 25$]{\includegraphics[width=2.0in]
{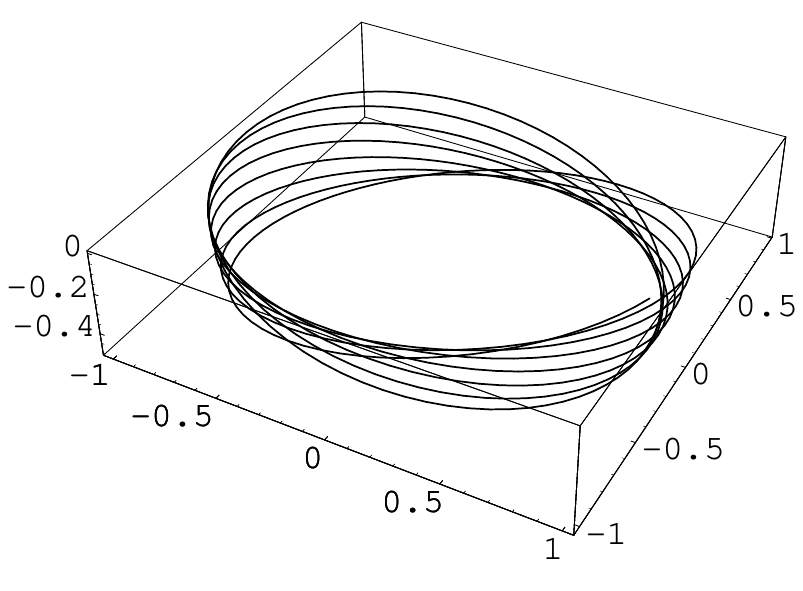}}
\hspace{0.5cm}
\subfigure[$\, \tau = 100$]{\includegraphics[width=2.0in]
{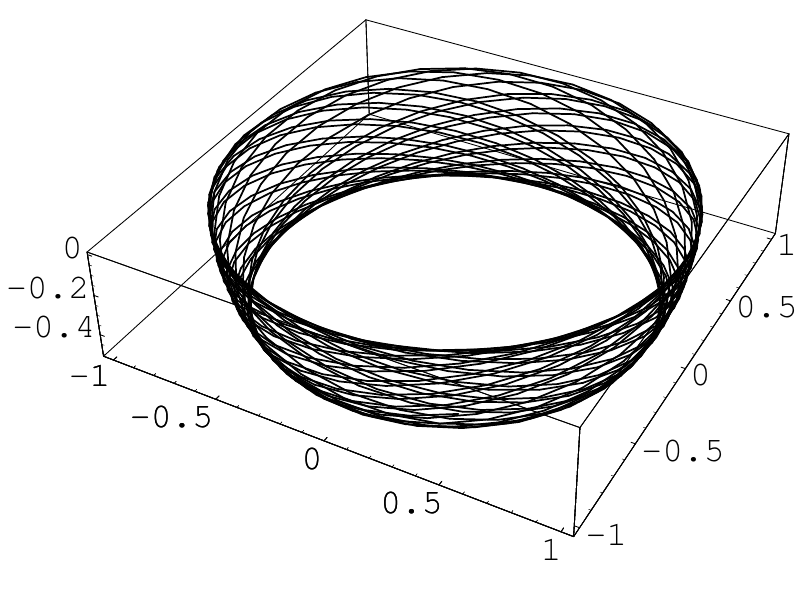} }
\caption{Trajectory for k=4, $\theta(0) = \pi/3$, $\theta^{'}(0) = 0$}
\label{traj2}
\end{figure} 
The period of oscillation can be computed as,
\beq
T = \frac{2}{\sqrt{k}} \int_{\alpha}^{\beta} \frac{d \theta}{\sqrt{
(\cos\alpha - \cos\theta)(\cos\theta - \cos\beta)}}
\eeq

The trajectory of the bead is confined within a band 
from $\theta =\alpha$ to $\theta =\beta$, 
on the surface of a sphere of radius $a$ (Figure \ref{traj2}). \\

\noindent
(iii) $\varepsilon = -1$ : \\
The motion of the bead is confined within 
$\theta = 0$ and a certain $ \Omega_2$, where ${\theta^{'}}$ is zero and
$U$ matches $\varepsilon$ (see Figure \ref{epot2}). However, the motion
is not oscillatory. If the bead were to start at $\Omega_2$,
$\theta^{'}(\Omega_2)=0$, however, $\theta^{''}(\Omega_2)< 0$, as
$\Omega_2 > \Omega_1$. 
As a result, $\theta^{'}$ becomes negative and hence $\theta$ decreases 
towards $\Omega_1$. At $\Omega_1$, $\theta^{''} =0$, whereas $\theta^{'}$ 
has maximum magnitude but is still negative.
\beq
|\theta^{'}_{max}| = |\theta^{'}_{\theta = \Omega_1}| = 
\sqrt{k} - \frac{1}{\sqrt{k}}
\eeq
Therefore $\theta$ continues to decrease towards zero.
At $\theta = 0$, $\theta^{'}=0$ and $\theta^{''}$ is zero as well. 
Depending on its initial position, the bead will eventually 
reach $\theta = 0$ , either directly, or after turning at 
$ \pm \Omega_2$, with ever decreasing speeds.
Once at this position, the bead will remain 
there until perturbed.
The period of oscillation of the bead may be computed by direct integration
as was done for the case of $0 \le k < 1$ with $\varepsilon = 1$.
However, by invoking the time reversal symmetry of the system,
one can show that the bead will take an infinite amount of time to
reach $\theta=0$. Let us start with the premise that the bead reaches 
$\theta = 0$ after a finite time $\tau_1$. At $\tau_1$, $\theta^{'}=0$,
and we may equally say that at $\tau = \tau_1$, the direction of 
$\theta^{'}$ has been reversed. 
Due to time reversal symmetry, the bead must then retrace its path.
Having reached $\theta=0$ however, the bead must stay there forever,
as $\theta^{'}$ and $\theta^{''}$ are both zero at this position.
Hence the bead cannot reach $\theta=0$ in finite time.\\

\noindent
(iv) $-1 < \varepsilon < 1$ : \\
The bead oscillates between $\theta = \pm \Lambda$ corresponding to
the turning points $D$ and $D'$ in Figure \ref{epot2}, where
$\theta^{'}$ is zero. Here, the bead has enough energy to overcome
the local maxima at $0$ and cross over to the other side. 
In a single period of oscillation, 
$|\theta^{'}|$ attains its maximum value four times, as the bead 
crosses both the minima $\Omega_1$ and $-\Omega_1$ in each half
oscillation.
The time period can be calculated as,
\beq
T = 4 \int_{0}^{\Lambda} \frac{d \theta}{\sqrt{(\cos\theta - 
\cos\Lambda)( 2 - k(\cos\theta + \cos\Lambda))}} 
\eeq

\noindent
(v) $\varepsilon =1$ : \\
Both $\theta^{'}$ and $\theta^{''}$ are zero at $\theta = \pm \pi$. 
Hence, if the bead were at
an initial position $-\pi < \theta < \pi$, it will eventually 
reach $\pm \pi$ at ever decreasing speeds. 
The speed, $|\theta^{'}|$ will be maximum at $\pm \Omega_1$ and will 
have a local minimum at $\theta=0$. \\

\noindent
(vi) $\varepsilon > 1$ : \\
$\theta^{'}$ is never zero as $\varepsilon$ is greater than $U(\theta)$
for all $\theta$. Thus $\theta$ increases or decreases indefinitely, depending on the sign of $\theta^{'}$. The bead will whirl around the entire 
hoop, in the same direction, past its initial position, periodically. 
The speed of the bead will increase and decrease periodically, 
with a local minimum at $\theta=0$, global minimum at $\pm \pi$ 
and maximum at $\pm \Omega_1$. The trajectory will be similar to the
ones shown in Figure \ref{full1}.

\section{Energy and Constraint forces}
\label{energy}
From (\ref{kepe}) the total energy of the bead is,
\beq
E = \frac{1}{2} ma^2\Big[\dot\theta^2 + \omega^2 \sin^2\theta - \frac{2g}{a}\cos\theta\Big] \; .
\eeq
Recalling the expression for the first integral of motion in (\ref{effe}),
we may write,
\beq
E = mga \cdot \varepsilon + ma^2\omega^2\sin^2\theta \; .
\eeq
The rate of change of energy is given by,
\beq
\frac{dE}{dt} = ma^2\omega^2 \dot\theta \sin2\theta \; ,
\eeq
$E$ is an even function of $\theta$. This implies that the 
net change in energy over a path that takes the bead from $\theta$ to 
$-\theta$ or back to $\theta$, is zero, although $\dot\theta$ may not be
zero along the path.
This accounts for the different oscillatory motions of the bead 
about $0$ described in section \ref{trajectory}.

The constraint forces exerted by the hoop may be calculated using 
the method of Lagrange's undetermined multipliers. Introducing two 
Lagrange multipliers,
$\lambda$ and $\mu$, the Lagrangian of the system may be written as,
\beq
L = \frac{m}{2}\big[\dot r^2 + r^2(\dot\theta^2 + \sin^2\theta \dot\phi^2)\big] + mgr\cos\theta + \lambda(r - a) + \mu(\phi - \omega t) \; ,
\eeq
with the constraint equations
\beq
r = a \;\;\;\; {\mbox{and}} \;\;\; \phi = \omega t \; .
\eeq
Using the Euler-Lagrange equation, the equations of motion for
$r, \phi $ and $\theta$ are,
\beqa
&& \ddot r  = r\big[\dot\theta^2 + \dot\phi^2 \sin^2\theta \big] + \frac{\lambda}{m} + g \cos\theta \; ,\\
&& \frac{d}{dt} (mr^2  \dot\phi \sin^2\theta) = \mu \; , \\
&& \frac{d}{dt} (r^2 \dot\theta) = r^2 \dot\phi^2 \sin\theta \cos\theta - gr\sin\theta \; .
\eeqa
Substitution of the constraint equations in the above yield,
\beqa
\lambda &=& -ma\big[\dot\theta^2 + \omega^2\sin^2\theta \big] - 
mg\cos\theta \; , \\
\mu &=& ma^2 \omega \dot\theta \sin2\theta  \; .
\eeqa
$\lambda$ and $\mu$ are the constraint force and torque along the 
radial and azimuthal directions respectively \cite{goldstein}.
As the hoop is assumed to be frictionless, it can only exert forces 
in a plane perpendicular to it. Hence, the $\theta$ component of the 
constraint force must be zero. This can be verified by the Newtonian
approach. Let $F_\theta$ denote the constraint force along 
the $\hat\theta$ direction. Balancing the forces along $\hat{\theta}$ 
direction we get
\beqa
&& F_\theta - mg\sin\theta = ma \big(\ddot\theta - \omega^2\sin\theta \cos\theta \big) \; , \; \mbox{or,} \\
&& F_\theta = ma\big[\ddot\theta + \frac{g}{a} \sin\theta - \omega^2 \sin\theta \cos\theta \big] \; .
\eeqa
The factor within the brackets is zero as per (\ref{lag2}), hence
$F_\theta = 0$. 

Therefore, the rate of work done by the constraint forces is,
\beq
P = F_\phi (a \omega\sin\theta) = \mu \omega =
 ma^2\omega^2\dot\theta \sin2\theta \; .
\eeq 
which is equal to the rate of change of energy of the bead.

\section{Phase Portraits and Bifurcations}
\label{phasebifur}
We now study the nature of fixed points and bifurcations, and portray the
phase space trajectories of the bead-hoop system.
By defining a new variable $\theta_1 = \theta^{'}$, (\ref{eom}) may be 
transformed to a set of two first order equations as,
\beq
\label{ord1}
\theta^{'} = \theta_1  \; , \;\; {\mbox{and}} \;\; 
\theta^{'}_1  = -\sin \theta (1- k \cos \theta) \; .
\eeq
The system described by (\ref{eom}) is conservative as there exists a 
first integral of motion which was computed in section \ref{physys}.
This conserved quantity represents an energy surface in the phase plane. 
The system motion takes place on this surface at constant height. 
That is, the effective energy remains a constant on the trajectories
of the system. Therefore, any isolated minimum or maximum of this 
energy surface is a center and there cannot be any attracting fixed points
or limit cycles\cite{strogatz}.
The system is also reversible, as under the transformations,
\beq
(i) \, t \to -t, \;\;\;  \theta \to -\theta \; , \;\; {\mbox{and}} \;\;
(ii) \, t \to -t, \;\;\;  \theta_1 \to -\theta_1  
\eeq
the governing equations (\ref{ord1}) remain invariant.
This means that if $(\theta(t), \theta_1(t))$ is a solution, so are
$(-\theta(-t), \theta_1(-t))$ and $(\theta(-t), -\theta_1(-t))$. 
The reversibility of the system does not admit any repelling fixed 
points. This can
be argued as follows. Let us assume that a repelling fixed point does
exist. If this is not at the origin, then by virtue of reversibility of
the system, there must exist an attracting fixed point, located at the
reflection of the repelling fixed point about $\theta$ or $\theta_1$
axis. This however, is not possible as the system is conservative.

Were the origin to be a repeller, then trajectories infinitesimally close
to the origin will move away from it in all directions. Reversibility requires
that for every outward bound trajectory close to the origin in the 
first quadrant, there exists a trajectory moving towards the origin in the
fourth quadrant. This is in contradiction to the assumption that the origin
is a repeller.
Hence, the fixed points of the bead-hoop system are either centers 
or saddles.

Equations (\ref{ord1}) show that for all values of
$k$, the fixed points
lie on the $\theta$ axis, and in the first quadrant, all trajectories
move right, as $\theta^{'} > 0$.

When $0 \le k \le 1$, there are two fixed
points for $\theta \, \in [0,\pi]$. When $ k > 1$, an additional fixed 
point appears at $\theta = \cos ^{-1}(1/k)$
We first investigate the fixed points and their nature by 
linear stability analysis. 
The effect of nonlinear terms is considered
subsequently. 
A general form of (\ref{ord1}) may be written as,
\bd
\theta^{'} = f(\theta,\theta_1) \;, \;\;\;
\theta_1^{'}= g(\theta,\theta_1)  \; .
\ed
A Taylor expansion about a fixed point $(\theta^{*}, \theta_1^{*})$ yields,
\beqar
\xi^{'} &=& f(\theta^{*},\theta_1^{*})+
\xi \frac{\partial f}{\partial \theta}\Big \vert_{\theta^{*}, \theta_1^{*}}
+ \eta \frac{\partial f}{\partial \theta_1}\Big 
\vert_{\theta^{*}, \theta_1^{*}} +
O(\xi^2,\eta^2,\xi\eta) +\dots \\ 
\eta^{'} &=& g(\theta^{*},\theta_1^{*})+
\xi \frac{\partial g}{\partial \theta}\Big \vert_{\theta^{*}, \theta_1^{*}}
+ \eta \frac{\partial g}{\partial \theta_1}\Big \vert_{\theta^{*}, 
\theta_1^{*}} +
O(\xi^2,\eta^2,\xi\eta) +\dots 
\eeqar
where, $\xi = \theta-\theta^{*}$ and $\eta = \theta_1-\theta_1^{*}$.
For small perturbations about the fixed point $(\theta^{*}, \theta_1^{*})$,
quadratic and higher terms in $\xi$ and $\eta$ may be neglected.
The equations governing $\xi$ and $\eta$ are given by,
\beq
\label{jac0}
\begin{pmatrix}
\xi^{'} \\
\eta^{'} 
\end{pmatrix}
=
\left(
\begin{array}{ccc}
\frac{\partial f}{\partial \theta} & \;\;\;
\frac{\partial f}{\partial \theta_1} \\
\frac{\partial g}{\partial \theta} & \;\;\; 
\frac{\partial g}{\partial \theta_1} 
\end{array} \right)_{\theta^{*}, \theta_1^{*}}  \; .
\eeq
The matrix on the rhs is the Jacobian matrix at the fixed point
$(\theta^{*}, \theta_1^{*})$.

The Jacobian matrix at the fixed point $(0,0)$ is, 
\beq
\mathbf{J}(0) = \left(
\begin{array}{ccc}
0 & \;\;\;1 \\
k - 1 & \;\;\;0 
\end{array} \right)  \; .
\eeq
As $k - 1 < 0$, the eigenvalues $\lambda = \pm i\sqrt{1 - k}$ are 
purely imaginary. Thus $(0,0)$ is a linear center with angular 
frequency of revolution given by $\omega \approx \sqrt{1 - k}$.
On expanding $\varepsilon$ in the 
neighbourhood of the origin, 
\beq
2 \varepsilon = \theta_1^2 + (1 - k) \theta^2 -1 + {\mbox{O}}(\theta^4) \; ,
\eeq
it is clear that $(0,0)$ is a minimum of the effective energy.
We therefore conclude that $(0,0)$ is 
a non-linear center as well.

Thus, the trajectories close to the origin are elliptical with eccentricity 
$e = \sqrt{k}$. The trajectories start as circles when $k = 0$, and become 
elongated in the $\theta$ direction as $k$ increases. 
For $(\pi,0)$, the Jacobian matrix is,
\beq
\mathbf{J}(\pi) = \left(
\begin{array}{ccc}
0 & \;\;\;1 \\
k + 1 & \;\;\;0 
\end{array} \right)  \; .
\eeq
The eigenvalues of this matrix are real with opposite signs, 
$\lambda = \pm\sqrt{k + 1}$. 
Hence, $(\pi,0)$ is a saddle with eigenvectors given by,
$\mathbf{v} = (1\;\;\lambda)^T$. Therefore, $(1\;\;\sqrt{k + 1})^T$ 
is the unstable manifold with $\lambda = \sqrt{k+1}$ and 
$(1\;\;-\sqrt{k + 1})^T$ is the stable manifold with $\lambda = -\sqrt{k+1}$.
They have equal and opposite slopes, whose value increases from 1 to 
$\sqrt{2}$ as $k$ increases from $0$ to $1$.

\begin{figure}[h]
\centering
\subfigure[$ \, k=3/4$]{\includegraphics[width=6.5cm]
{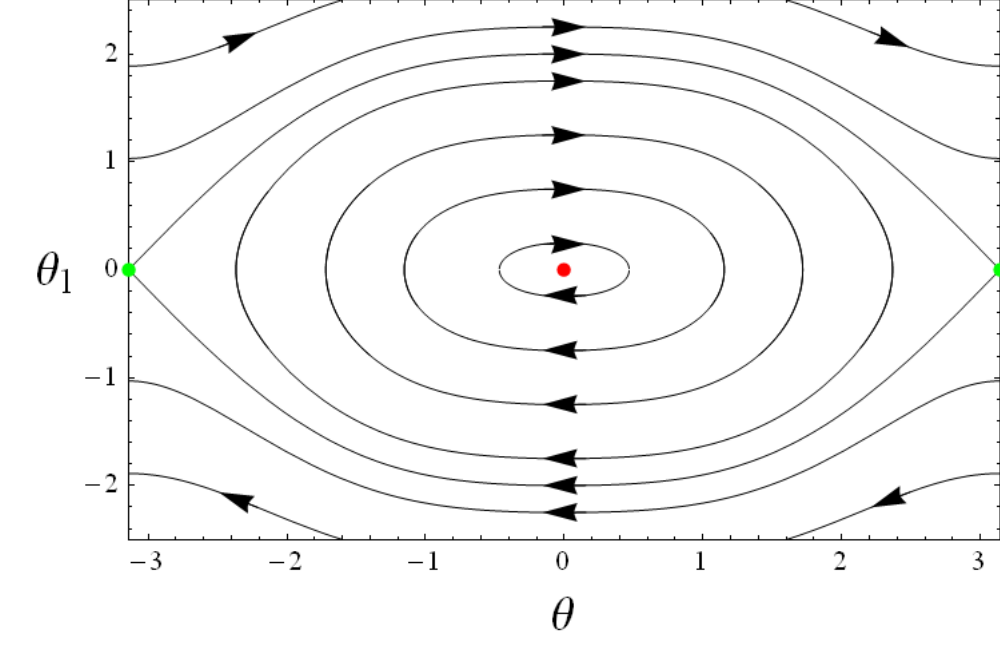} }
\hspace{1cm}
\subfigure[$\, k=1$]{\includegraphics[width=6.5cm]
{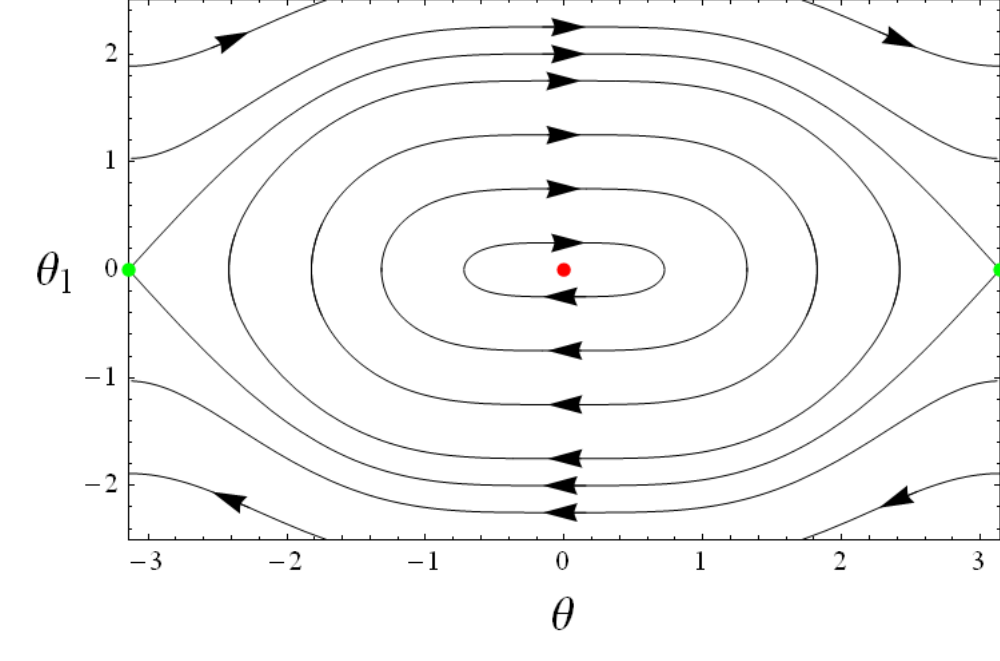}}
\caption{Phase portraits for different $k$ values}
\label{port1}
\end{figure}

The phase trajectories for $k=0.75$ and $k = 1.0$ are shown in Figure
\ref{port1}. 
One can identify heteroclinic orbit or saddle connection, which is a 
trajectory connecting two saddles. 
A trajectory starting infinitesimally close to $\theta = -\pi$ with 
$\theta_1 = 0$, crosses the $\theta_1$ axis at $\sqrt{2(\varepsilon + 1)} 
$. By reversibility, this trajectory must reach $\theta = \pi$, thus 
forming a heteroclinic orbit. It may be noted here that $\theta = \pi$
and $\theta = -\pi$ correspond to the same position on the hoop. The
$\theta_1 - \theta$ plane is like the curved surface of a cylinder, with the
$-\pi$ and $\pi$ edges joined. The heteroclinic orbits in Figure \ref{port1}
are thus homoclinic orbits on the cylinder starting and ending on the
saddle at $(\pi,0)$.
The closed orbits around the center at origin represent 
periodic oscillations about $\theta = 0$, for $-1 < \varepsilon < 1$,
traditionally called librations. The saddle connection or heteroclinic
orbit represents the delicate motion of the bead where it starts near 
the top of the hoop, swings past the bottom of the hoop and slows to a 
halt as it approaches the top of the hoop again. 
The trajectories above the saddle connection correspond to the 
whirling motion or complete revolution of the bead around the hoop.\\

When $k=1$, the Jacobian at $(0,0)$ becomes,
\beq
\mathbf{J}(0) = \left(
\begin{array}{ccc}
0 & \;\;\;1 \\
0 & \;\;\;0 
\end{array} \right) \; .
\eeq
Consequently, both the eigenvalues vanish. This does not give us much 
information, so we expand the effective energy, $\varepsilon$, in 
the neighbourhood of the origin,
\beq
\varepsilon = \frac{\theta_1^2}{2} + \frac{\theta^4}{8} - 1 + 
{\mbox{O}}(\theta^6) \; .
\eeq
The point $(0,0)$ remains a minimum of the energy surface and hence is still 
a center, though very weak. The weakness of the center is brought out by 
the elongated shape of the orbits in the phase plane (Figure \ref{port1}). 
Close to $\theta =0$, the decay of $\theta$ is very slow.
This lethargic decay is termed `critical slowing down' and 
sometimes signifies the onset of bifurcation\cite{strogatz}. \\

When $ k > 1$, a new fixed point appears
at $(\Omega_1,0)$. As we noted in section \ref{trajectory}, 
the minima of $U(\theta)$
and hence the minima of the energy surface occur at $\pm\Omega_1$. 
Hence from our previous reasoning, $(\Omega_1,0)$ is a center. 
This can be further verified by forming the Jacobian at $(\Omega_1,0)$,
\beq
\mathbf{J}(\Omega_1) = \left(
\begin{array}{ccc}
0 & \;\;\;1 \\
-(k - 1/k) & \;\;\;0 
\end{array} \right)  \; .
\eeq
$\mathbf{J}(\Omega_1)$ has purely imaginary eigenvalues $\lambda = 
\pm i\sqrt{k - 1/k}$. Thus, $(\Omega_1,0)$ is a center, with angular 
frequency of libration $\omega \approx \sqrt{k - 1/k}$. 
Let $k = 1 + \delta$. Then $(1 + \delta)^{-1} = \cos\Omega_1$. 
Expanding both sides for small $\delta$ and equating the dominant terms, 
one finds that $\Omega_1 \sim {\mbox{O}}(\sqrt{\delta})$. 
Thus, $\Omega_1$ increases much more rapidly than $\delta$ does.

On the other hand, at $(0,0)$,
\beq
\mathbf{J}(0) = \left(
\begin{array}{ccc}
0 & \;\;\;1 \\
k - 1 & \;\;\;0 
\end{array} \right)
\eeq
now has real eigenvalues $\lambda = \pm\sqrt{k - 1}$. Therefore, 
the origin has transformed into a saddle having $(1\;\;\sqrt{k - 1})^T$ 
as the unstable manifold and $(1\;\;-\sqrt{k - 1})^T$ as the 
stable manifold. They become steeper as $k$ increases. The fixed point 
at $(\pi,0)$ continues to be a saddle as before.

\begin{figure}[h]
\centering
\subfigure[$\, k=1.1$]{\includegraphics[width=2.0in]{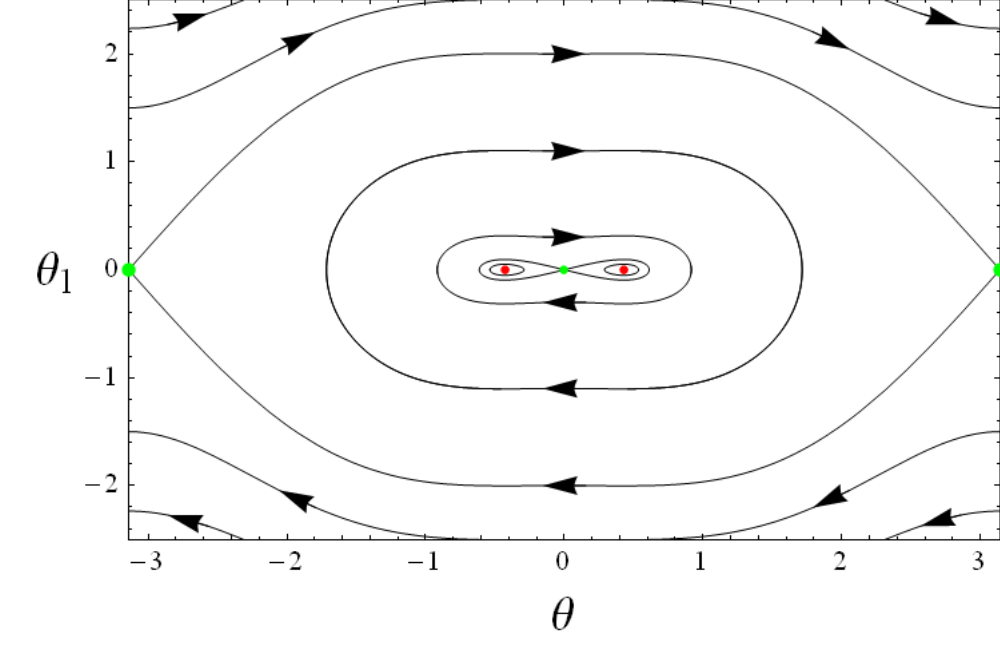}}
\subfigure[$\, k=1.1 (\mbox{central part})$]{\includegraphics[width=2.0in]
{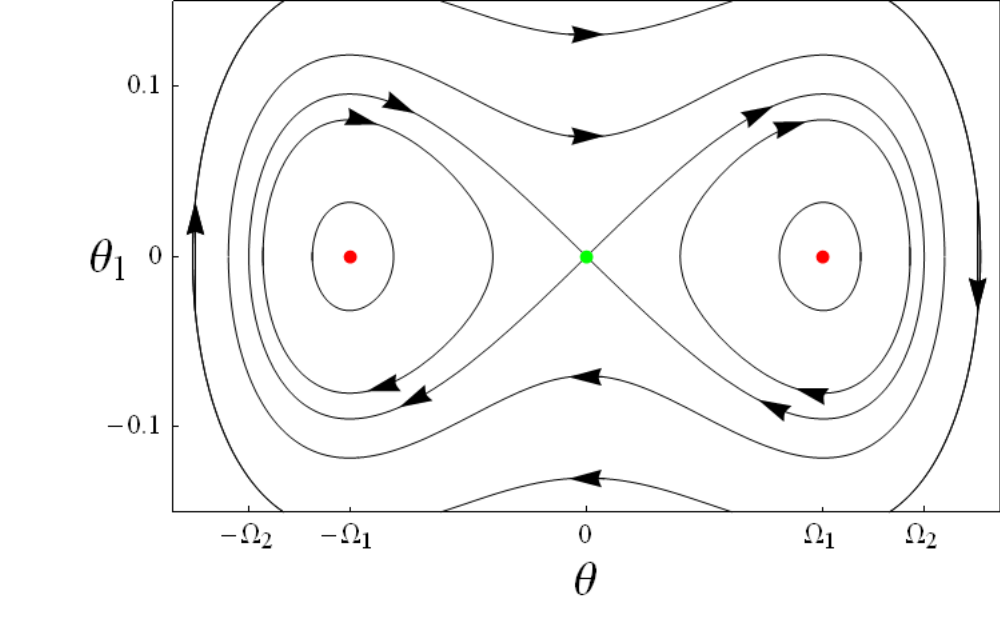} }
\subfigure[$\, k=4$]{\includegraphics[width=2.0in]
{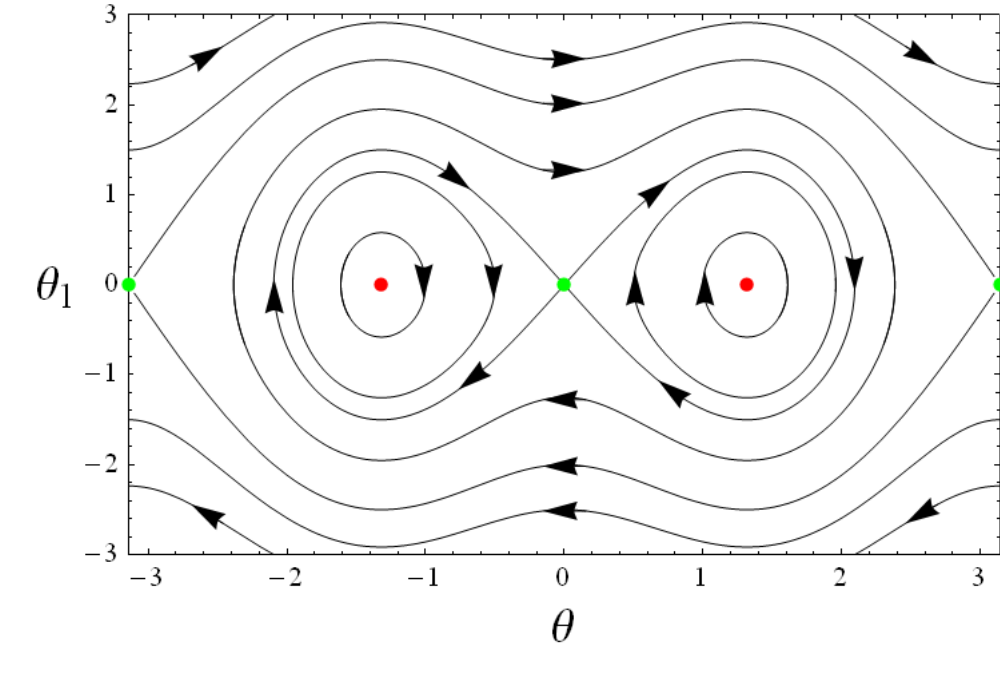}} 
\caption{Phase portraits for $k>1$ showing homoclinic orbits.}
\label{port2} 
\end{figure}

A trajectory starting at $\theta \to 0^+$ and $\theta_1 = 0$ cuts the 
$\theta$ axes exactly at $\theta = \Omega_2$. By reversibility, we 
conclude that this forms a homoclinic orbit, enclosing the center at 
$(\Omega_1,0)$. The heteroclinic orbits remain as before. 
We get closed orbits in the region of the phase-plane bounded by 
the homoclinic and the heteroclinic orbits (Figure \ref{port2}). 
By expanding the effective energy about $(\Omega_1,0)$, we find that 
the orbits near the center are approximate ellipses described by,
\beq
\theta_1^2 + \Big(k-\frac{1}{k}\Big)(\theta - \Omega_1)^2 = 
2\varepsilon + \Big(k-\frac{1}{k}\Big) + {\mbox{O}}((\theta - \Omega_1)^3)
\eeq 
with eccentricity $e \approx \sqrt{|k-1/k -1|}$. These features are 
shown in the phase portraits in Figure \ref{port2}.
The physical interpretation of the trajectories is as follows : the fixed point $(\Omega_1,0)$ is the stable equilibrium position. The small orbits around it represent periodic oscillations or librations. One can identify homoclinic 
orbits which begin and end at the same fixed point, the saddle at $(0,0)$.
This corresponds to the motion of the bead where it slows to a halt at 
$\theta = 0$. The closed orbits surrounding these correspond to 
oscillations about $\theta = 0$ with amplitude greater than $\Omega_2$ 
of Figure \ref{epot2} in section \ref{trajectory}. The heteroclinic orbits 
indicate the delicate motion where the bead slows to a halt at the top of 
the hoop. The trajectories outside it represent periodic whirling or complete
revolutions of the bead around the hoop. 
These inferences are consistent with the motion of the bead as discussed
in section \ref{trajectory} using graphical and differential equation 
analysis.

A change in the number and nature of fixed points
of a dynamical system, as one of its parameters is varied is called
bifurcation. When $k > 1$, the fixed point at $(0,0)$ changes from
a center to a saddle. 
Two new centers emerge on both sides of $(0,0)$ at $\pm \cos^{-1}(1/k)$, 
moving away from the origin as $k$ increases further. 
Therefore, the system undergoes a supercritical pitchfork bifurcation 
at the origin as $k$ is varied through $1$ \cite{strogatz}. 
Also, $|\mathbf{J}(0)| = 0$ for $k = 1$. So, it can be called a 
zero eigenvalue bifurcation. It is also referred to as
a symmetry-breaking bifurcation. 
\begin{figure}[h]
\centering
\includegraphics[width=10cm]{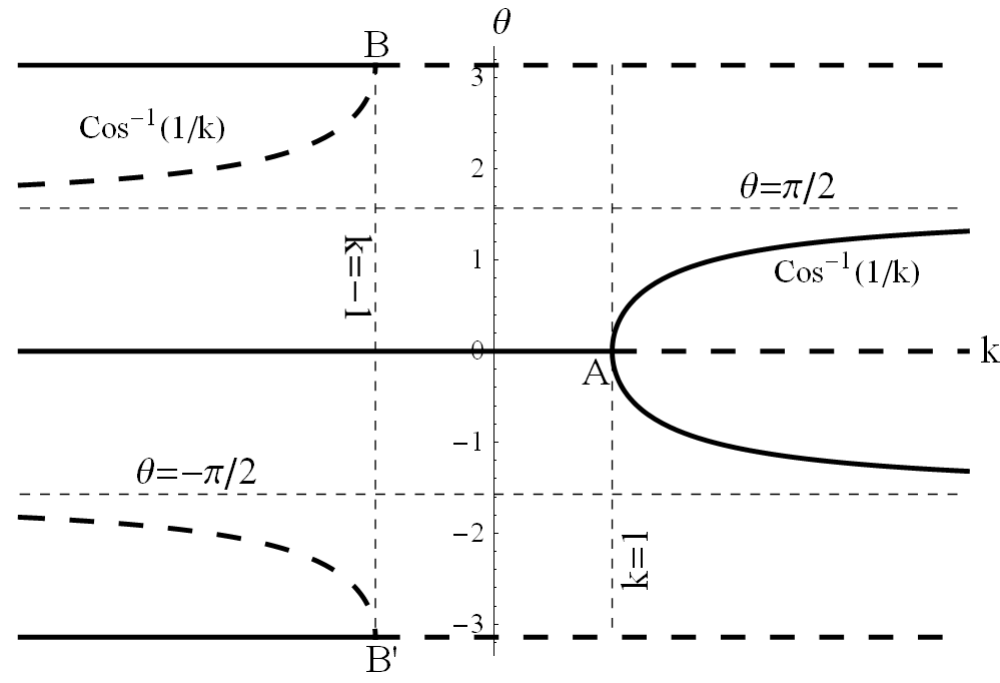}
\caption{The bifurcation diagram for both positive and negative $k$. 
Pitchfork bifurcation takes place at $A$ and $B$. $B$
and $B'$ represent the same point.}
\label{bifur}
\end{figure} 
The bifurcation diagram in Figure \ref{bifur}, encompasses both 
positive and negative values of $k$. For the bead-hoop system that we have
studied, negative values of $k$ have no physical meaning. However,
there are systems with governing equations similar to (\ref{ord1}),
where negative values of $k$ are allowed. For example, a charged particle moving on a vertically rotating circular wire, with a uniform 
magnetic field $B_0$ in the downward vertical direction, has the same 
governing equations as (\ref{ord1}), but with the parameter
$k$ given by,
\beq
k = \frac{\omega^2 }{(g/a)} - \frac{qa\omega B_0}{mg}
\eeq
This allows negative values of $k \geq -\omega_L^2/\omega^2_c$, where 
$\omega_L$ is the Larmor frequency. For this system, negative $k$ 
values become relevant.
In Figure \ref{bifur}, the solid lines denote center 
and the dashed lines denote saddle. As $k$ is made more negative than -1, 
$(\pm\pi,0)$ changes from a saddle to a center and two symmetrical saddle points fork out towards $\theta =\pm \cos^{-1}(1/k)$ 
(pitchfork / symmetry-breaking bifurcation). If we imagine the diagram to be on a cylindrical surface with the top and bottom borders joined end to end, then the figure is very symmetrical (Figure \ref{bifur}). 

\section{Connections with other systems}
\label{similar}
The system we have studied has interesting connections and similarities with
a variety of other physical systems. As mentioned previously, the same set of 
equations are obtained for a charged particle moving on a vertically 
rotating circular wire, in the presence of a uniform magnetic field directed
vertically downward. However, the parameter $k$ can take negative values as well. The
phase portraits, nature of fixed points and bifurcations can thus be analyzed
in a similar manner.

The system exhibits spontaneous symmetry breaking at $k=1$ 
($\omega = \omega_c$), quite similar to a Landau second-order
phase transition\cite{sivardiere, landau, fletcher, mancuso}.
The critical angular velocity of the hoop, $\omega_c$, is analogous to
the critical temperature $T_c$ and the equilibrium position of the bead to
the order parameter in a Landau system.

Another example is the Duffing oscillator described by the equation,
\beq
\ddot x + x + \epsilon x^3 = 0 \; .
\label{duff}
\eeq
Expanding (\ref{eom}), for $0 \leq k \leq 1$, in the neighbourhood 
of $\theta = 0$ we get,
\beq
\theta^{''} + (1 - k) \theta + \left(\frac{2k}{3}-\frac{1}{6}
\right)\theta^3 
= {\mbox{O}}(\theta^5) \; .
\eeq
If we define $u=\sqrt{1-k} \, \tau$, then the above equation may be
written as,
\beq
\frac{d^2 \theta}{d u^2} + \theta + \frac{4k-1}{6(1-k)}\theta^3 =
{\mbox{O}}(\theta^5) \; ,\qquad \mbox{for $0 \leq k < 1$}.
\eeq
Neglecting terms of order $5$ and higher in 
$\theta$, the equation reduces to the Duffing oscillator equation 
(\ref{duff}). Thus the nature of fixed points and bifurcations studied 
for the bead-hoop system in the small amplitude limit, with $ 0 \le k < 1$,
hold for the Duffing oscillator as well. 
The Duffing equation (\ref{duff}) describes the undamped motion of a 
unit mass attached to a nonlinear spring with restoring force 
$F(x) = -x - \epsilon x^2$. The Duffing equation is also a conservative 
system. The coefficient $\epsilon$, of the nonlinear term is related to the
parameter $k$ of the bead-hoop system as $\epsilon = (4k -1)/(6(1-k))$. As $k$ 
is varied from $0$ to $1$, $\epsilon$ increases from $-1/6$ to $\infty$.
For this range of $\epsilon$, the Duffing oscillator does indeed have a 
nonlinear center at the origin\cite{strogatz}.

A variation of our system is the hoop rotating about a horizontal axis. This 
system can operate as a one-dimensional ponderomotive particle 
trap\cite{johnrab}. 

The rigid pendulum can be considered a special case of our system 
with $k = 0$, and many problems in various branches of physics, such as,
the theory of solitons, the problem of superradiation in quantum optics, 
and Josephson effects in weak superconductivity can be reduced to 
the differential equation describing the motion of a pendulum\cite{butikov}.

Let us now discuss briefly the case when the hoop is not maintained at constant
rotation, i,e., $\dot\phi$ is not a constant. Initially, the hoop 
is given some non-zero $\dot\phi$ and the bead-hoop system is left to itself.
The total energy of the system is conserved, but both $\dot\theta$ and 
$\dot\phi$ vary with time. From the Lagrangian of the system, 
one can arrive at the following,
\beqar
\dot\phi &=& \frac{k_1}{M + m \sin^2\theta}  \; ,\\
\theta^{''} &=& - \sin\theta \Big[1 - \frac{a_1 \cos\theta}{(1 + b \sin^2\theta)^2} \Big] \; ,
\eeqar 
where $M$ and $m$ denote the masses of the hoop and the bead respectively, and $a_1 = k_1^2 a/(M^2 g)$, $b = m/M$. When $b \ll 1$, i.e., $m \ll M$,
$\dot\phi \approx \frac{k_1}{M} = \omega$, a constant.
In this limit, the hoop has very large inertia compared to the bead, 
and therefore a high tendency to resist any change in its energy. 
\beqa
a_1 &=& \left(\frac{k_1}{M}\right)^2\frac{a}{g} = 
\frac{\omega^2 }{\omega^2_c} \; ,\\
{\mbox{Hence,}} \qquad
\theta^{''} &=& - \sin\theta \left(1 - \frac{\omega^2 
a}{g} \cos \theta \right) \; .
\eeqa 
which is the same as (\ref{eom}).

\section*{Conclusion}
The diverse modes of motion of a bead moving without friction, on a 
vertically rotating circular hoop have been explored using a simple 
theoretic approach based on symmetry arguments and elementary calculus. 
This simple system exhibits
several features of nonlinear dynamics and hence this study can serve as
a good background for investigating more complicated nonlinear systems.

\end{document}